\documentclass{aa}  

\usepackage{graphicx}
\usepackage{amsbsy}
\usepackage{txfonts}
\usepackage{pdfpages}
\usepackage{natbib}
\usepackage{amsmath}
\usepackage{amscd}
\usepackage{booktabs}
\usepackage{multirow}
\usepackage{subfig}
\usepackage{resizegather}
\usepackage{url}
\usepackage{bm}
\usepackage{lineno}
\bibpunct{(}{)}{;}{a}{}{,} 

\begin{document}

   \title{The seven sisters DANCe III.}
   \subtitle{Projected spatial distribution}
   \author{Olivares, J.
          \inst{1,2,3}
          \and
          Moraux, E.
          \inst{2}
          \and
          Sarro. L.M.
          \inst{1}
          \and
          Bouy, H.
          \inst{3,4}
          Berihuete, A.
          \inst{5}
          \and
          Barrado,D.
          \inst{4}
          \and
           Huelamo, N.
          \inst{4}
          \and
          Bertin, E.
          \inst{6}
          \and
          Bouvier, J.
          \inst{2}
          }

   \institute{Dpt. de Inteligencia Artificial , UNED, Juan del Rosal,
     16, 28040 Madrid, Spain\\ \email{lsb@dia.uned.es} 
     \and
     Univ. Grenoble Alpes, CNRS, IPAG, F-38000, Grenoble, France\\
     \email{Javier.Olivares@univ-grenoble-alpes.fr}
     \and
     Laboratoire d'astrophysique de Bordeaux, Univ. Bordeaux, CNRS,
     B18N, all\'ee Geoffroy Saint-Hilaire, 33615 Pessac, France.
     \and
     Depto. Astrof\'{\i}sica, Centro de Astrobiolog\'{\i}a
     (INTA-CSIC), ESAC campus, Camino bajo del castillo s/n, 28692 Villanueva de la Cañada, Spain
     \and
     Dpt. Statistics and Operations Research, University of
     C\'adiz, Campus Universitario R\'io San Pedro s/n.  11510 Puerto
     Real, C\'adiz, Spain
     \and  
     Institut d’Astrophysique de Paris, CNRS UMR 7095 and UPMC, 98bis
     bd Arago, F-75014 Paris, France
   }

   \date{}

 
  \abstract
  {Membership analyses of the DANCe and Tycho+DANCe data sets provide the largest and least contaminated sample of Pleiades candidate members to date. }
   {We aim at reassessing the different proposals for the number surface density
      of the Pleiades in the light of the new and most complete list of
     candidate members, and inferring the parameters of the most
     adequate model.}
   {We compute the Bayesian evidence and Bayes Factors for
     variations of the classical radial models. These include elliptical symmetry, and luminosity segregation.
     As a by-product of the model comparison, we obtain posterior
     distributions for each set of
     model parameters.}
   {We find that the model comparison results depend on the spatial
     extent of the region used for the analysis.  For a circle of 11.5
     parsecs around the cluster centre (the most homogeneous and
     complete region), we find no compelling reason to abandon King's
     model, although the Generalised King model introduced here has slightly better fitting properties.  Furthermore, we
     find strong evidence against radially symmetric models when
     compared to the elliptic extensions. Finally, we find that
     including mass segregation in the form of luminosity segregation
     in the J band is strongly supported in all our models.}
   {We have put the question of the projected spatial distribution of
     the Pleiades cluster on a solid probabilistic framework, and
     inferred its properties using the most exhaustive and least
     contaminated list of Pleiades candidate members available to date. Our
       results suggest however that this sample may still lack about
       20\% of the expected number of cluster members. Therefore, this study
       should be revised when the completeness and homogeneity of the
       data can be extended beyond the 11.5 parsecs limit. Such a study will allow for more
     precise determination of the Pleiades spatial distribution, its tidal radius, ellipticity,
     number of objects and total mass.}  

   \maketitle
%

\section{Introduction}
\label{sec:intro}

The projected spatial distribution (PSD), also known as number surface
density, of a stellar cluster is the two dimensional (2D) projection, in
the plane of the sky, of its three dimensional (3D) space
distribution. Because celestial coordinates are far more easily
measured than parallaxes  (at least before \emph{Gaia}), only a
small fraction of the objects with stellar positions have distance
estimates. Furthermore, the relative uncertainties in the celestial
coordinates yield far more precise measurements (by a factor of
$10^4$) of distances perpendicular to the line of sight than those
achieved by parallaxes along this line so far (except perhaps for very
close objects). This explains why most of the previous works devoted
to studying the spatial distribution of stars in clusters have been done
using the PSD.

In the case of the Pleiades, cross-matching the \emph{Hipparcos} catalogue
\citep{1997A&A...323L..49P} with the 2109 candidate members of
\citet{2015A&A...577A.148B}, shows that only 70 of them have parallax measurements. 
This figure has roughly doubled with the first \emph{Gaia} data release
DR1 \citep{2016A&A...595A...1G}, and is expected to improve based on
the longer time baselines and hence more accurate measurements of
subsequent \emph{Gaia} releases. In preparation for the analysis of
these upcoming data sets and to narrow down the set of models
that will be tested in the context of 3D studies,
we have initiated a re-examination of the current analytical alternatives 
to describe the PSD of the Pleiades cluster.

The Pleiades PSD has been thoroughly studied in the past.
\citet{Pinfield1998} fitted King's \cite[][hereafter King's]{King1962}
empirical profiles to the positions of $1194$ candidate members from
the literature, which were contained in a $3^{\circ}$ radius area. For
their fitted King profiles, they used different mass ranges, with bins
centred at $5.2,1.65,0.83$ and $0.3 \,\mathrm{M_{\odot}}$. Using tidal
forces\footnote{We highlight that their Eqs. 10 and 12 seem to be slightly
  different from those reported \citet{2008gady.book.....B}.}, they
iteratively constrained the tidal radius to a value of $13.1$ pc
($\sim 5.6^{\circ}$). They infer a core radii in the 0.9-2.91 pc range
in the different mass bins, and a total mass of
$735\,\mathrm{M_{\odot}}$.  They interpreted the gradual increase in
the core radii for decreasing mass ranges as evidence of mass
segregation.

The same year, \citet{Raboud1998} also fitted a King's profile to a
list of 270 candidate members with masses in the range
$0.74-7.04\,\mathrm{M_{\odot}}$, which were contained within a
$5^{\circ}$ radius area. They found a core radius of $1.5$ pc and a
tidal radius of $17.5$ pc ($7.5$ degrees). Using different approaches,
they derived a total mass within the range of
$500 -8000 \,\mathrm{M_{\odot}}$. They also measured an ellipticity of
$\epsilon=0.17$. However, they did not make any explicit mention of
the position angle of the axis of the ellipse, and simply state that
it is roughly parallel to the galactic equator.

Later, \citet{Adams2001} also fitted a King's profile to objects with
membership probabilities $p>0.3$ within a radius of $10^{\circ}$. They
found a core radius of $2.35-3.0$ pc and a tidal radius of $13.6-16$
pc ($5.8 - 6.8^{\circ}$). They estimate a total mass of $\sim
800\,\mathrm{M_{\odot}}$, and their measured ellipticities are in the
range $0.1-0.35$.

\citet{Converse2008} fitted a King's profile to a sample of 1245
candidate members from the \citet{Stauffer2007} compilation. These
objects have masses greater than $0.08\,\mathrm{M_{\odot}}$ and are
contained within a $5^{\circ}$ radius. They obtained a tidal radius of
$18$ pc ($7.7^\circ$) and a core radius of $1.3$ pc.  They found
unambiguous evidence of mass segregation using a method they devised inspired by econometrics. Later, \citet{Converse2010} refined
their previous study \citep{Converse2008} and obtained a core radius
of $2.0\pm0.1$ pc, a tidal radius of $19.5 \pm 1.0 $ pc ($\sim 8.3$
degrees), a total number of systems of 1256$\pm$35, and a total mass
of $870\pm35\,\mathrm{M_{\odot}}$.

The previous summary of results shows at least two interesting
points. In the first place, the King's profile has been the preferred
choice for the Pleiades cluster, although it was created to fit the
PSD of globular clusters. Since globular clusters are farther away
than open clusters and in a low-density environment, the end
of their PSD is usually well within the survey area, which is not the case for
the Pleiades. The second point concerns the increasing trend of the
tidal radius with the size of the survey and the publication date (Table \ref{tab:tidal_iterature}); as the surveys increase in area, the
derived tidal radii increase as well. This may indicate that truncation
has not been accounted for (see Appendix \ref{app:truncation} and
Fig. \ref{fig:KingSynNoTrunc} particularly). The exception to this
trend is the work of \citet{Adams2001}, in which the tidal radius is
well within the survey radius.  Since these authors used low-membership-probability ($\geq0.3$) objects, their results may be
affected by a significant contamination rate, which these authors
acknowledge for their $>5^\circ$ sample.

The two points mentioned above are tightly related. With the
exception of the work of \citet{Adams2001}, the coverages of the rest
of the surveys have not reached their estimated tidal radius. This
indicates that the previously used samples of members were spatially
truncated. They only contain objects from the inner parts of the
cluster. Thus, estimates of the tidal radius may have been biased, and
were, in any case, highly correlated with the contamination rate.

\begin{table}[ht!]
\caption{Survey, and derived core and tidal radius for recent studies in the literature. }
\begin{center}
\begin{tabular}{lccc}
 &Core&Tidal & Survey\\
&radius&radius&radius\\
&(pc)&(pc/$^\circ$)&($^\circ$)\\
\hline
\hline
\citet{Pinfield1998}&0.9-2.91&13.1/5.6&3\\
\citet{Raboud1998}&1.5&17.5/7.5&5\\
\citet{Adams2001}&2.35-3.0&16/6.8&10\\
\citet{Converse2008}&1.3&18/7.7&5\\
\citet{Converse2010}&2.0&19.5/8.3&5\\
\hline
\end{tabular}
\end{center}
\label{tab:tidal_iterature}
\end{table}%

With \emph{Gaia} data coming up soon, we will have very
accurate measurements of the spatial distribution of all the brightest
($G\leq20$ mag) members of nearby clusters. Therefore, it is
important to define sufficiently complex models to describe these
measurements. The early and simple formulations of the PSD (e.g. King)
were perfect when a dozen or a few tens of dozens of members were
known. But the accuracy and completeness of future surveys will allow
us to look in finer detail.

The study of the spatial distribution also has implications that go beyond
its intrinsic interest. One of them is the existence of mass
segregation as a result of star formation and dynamical interactions
in the cluster. This effect has been predicted by numerical
simulations of the internal cluster dynamics; see, for example,
\citet{1987MNRAS.224..193T,2001MNRAS.321..699K,Moraux2004,Converse2010}.
Confirming and quantifying its dependance on various parameters
(e.g. initial mass function, core mass function, total mass of the
cluster, presence or absence or massive stars, T- or OB- association)
shall provide important input to the models and simulations of star
formation and dynamical evolution.

In the specific case of the Pleiades, mass segregation has been
reported in the works of
\citet{Raboud1998,Pinfield1998,2001MNRAS.321..699K,Adams2001,Moraux2004,Converse2008,Converse2010}. Yet, \cite{Loktin2006},
using radial and tangential velocity dispersions, found no hint of
mass segregation in a sample of 340 stars contained in the central
$2.3^\circ$. However, his results may arise from the low number and extent of his sample. All the
mentioned works performed their analyses by binning the stellar
samples in mass or distance ranges. It is well known however that
fitting a function to a binned data set can introduce biases
\citep{2003drea.book.....B,1989ApJ...342.1207N}, and that modifying
the bin width could improve the fitting to a preferred model
\citep{2012arXiv1209.2690T}. Thus, the use of bins in previous works
and the contradictory mass-segregation results found by
\cite{Loktin2006} may suggest that the hypothesis of mass segregation
in the Pleiades requires a more solid reexamination.

In this work we aim at addressing this hypothesis on the basis of the
largest and least contaminated sample of Pleiades candidate members
found to date: the combined list of candidate members from \citet{2015A&A...577A.148B} and \citet{Javier}. 
We avoid the binning biases by using
Bayesian inference methods applied to continuous and thus non-binned
distributions. In addition, these Bayesian methods allow a
quantitative comparison of the competing models, including those with
and without mass segregation. This will allow us to establish on firm
grounds the analytical expression of the Pleiades PSD, and its
potential dependence on stellar mass.

In Section \ref{sec:data} we briefly describe the data set that
forms the basis of our analysis. In Section \ref{sec:models} we
present the set of radially symmetric analytical models we used,
as well as their extension to biaxially symmetric (elliptical)
profiles. We also include a luminosity dependence of the core radius
(as a proxy to the investigation of mass segregation). In Section
\ref{sec:modelselect} we describe the foundations of model selection
in the Bayesian framework.  We then discuss and compare the results
that we obtain for the posterior distributions of the various models
in Section \ref{sec:results}, where we also briefly describe our estimates on the
total mass and number of members in the cluster. Finally, in Section
\ref{sec:conclusions} we summarise the conclusions drawn from the
study.


\section{The data sample}
\label{sec:data}

The data set used to compare the models in Section
\ref{sec:models} corresponds to the high-membership-probability candidate members of \cite{Javier}, in the middle and faint luminosity end,
with the addition of the Tycho-2 Pleiades high-luminosity candidate members from \cite{2015A&A...577A.148B}. This joint data set comprises the equatorial coordinates R.A. and Dec. (in the following $\alpha$ and $\delta$), proper motions, photometry, and membership probabilities of 2060 sources.  In this analysis we work only with the positions, membership probabilities, and J photometric band. The latter is the reddest most available band for this list of members, and is used as a proxy for the mass and to explore evidence of mass segregation.

\subsection{Completeness of the sample}
To properly establish the probabilistic framework, it is necessary to
take into account the observational constraints of the data. The
Pleiades DANCe catalogue is constrained by its sky coverage 
and the different degrees of completeness \citep[see][ for
details]{2013A&A...554A.101B,2015A&A...577A.148B}. Although the data set extends up to a radius of $6.5^{\circ}$, \citet{2015A&A...577A.148B} conservatively assume that the census is homogeneous in coverage and limiting magnitude only in the central $3^{\circ}$ radius area.

\begin{figure}[ht!]
\begin{center}
  \includegraphics[page=1,width=\columnwidth]{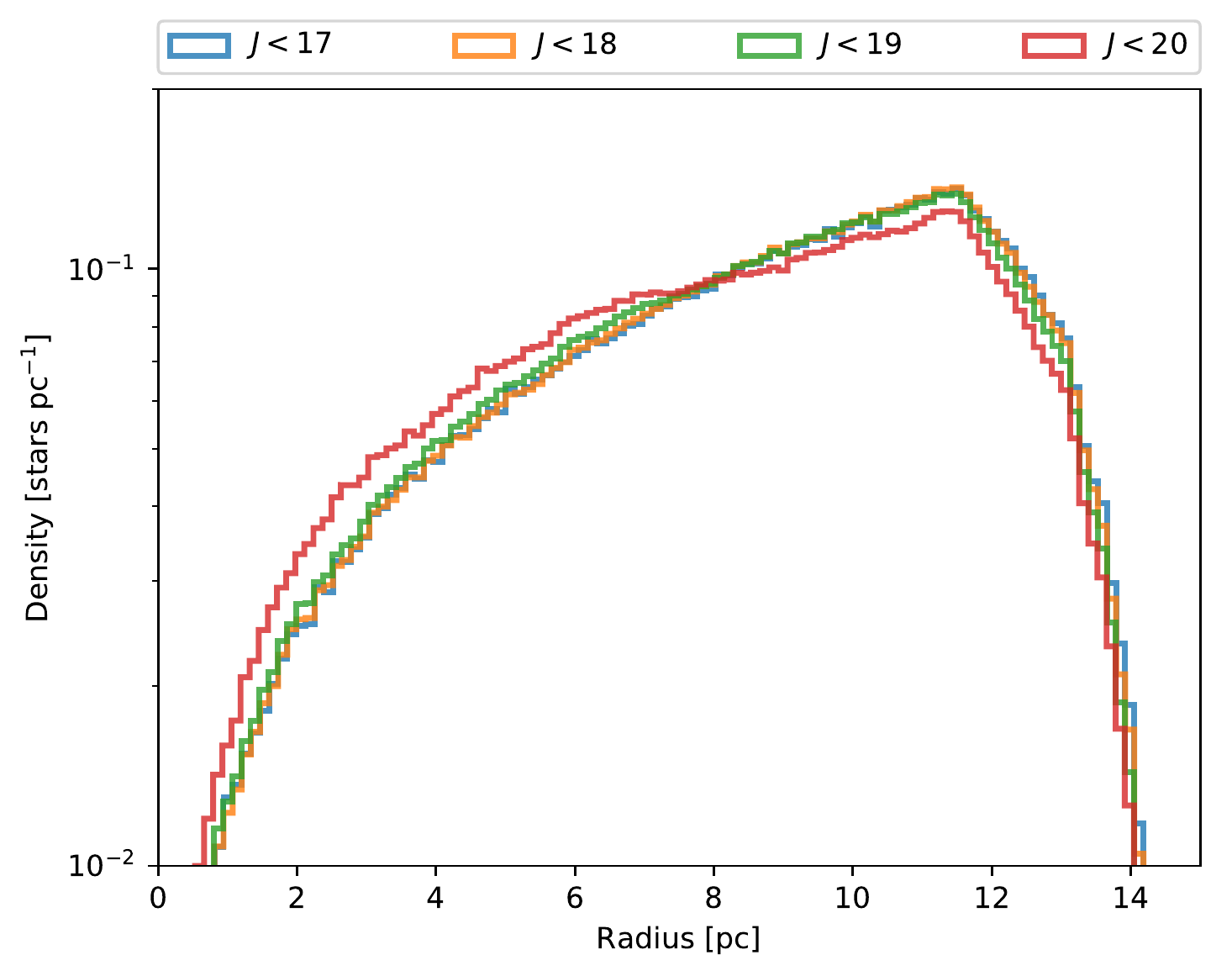}
  \includegraphics[page=2,width=\columnwidth]{Analysis/RadiiDistribution_Tycho+DANCe_Jmag.pdf}
\caption{Density of sources in the combined DANCe+Tycho catalogue as a function of the radial distance to the cluster centre and the J magnitude. Top panel: All sources contained within the limiting magnitudes. Bottom panel: Sources within the 11.5 pc radius of spatial completeness (vertical grey line), and binned in magnitudes. The black line represents the density of two million sources uniformly distributed in the plane of the sky.}
\label{fig:completeness}
\end{center}
\end{figure}

Here, we estimate the completeness of the whole of the joint Tycho+DANCe survey in terms of the J band luminosity and spatial coverage, which also applies to our list of candidate members. In Fig. \ref{fig:completeness} we show the distributions of the number of sources in the combined DANCe+Tycho catalogue as a function of the radial position for different limiting magnitudes and bins in the J band. The radial position is computed assuming a distance of 134.4 pc to the Pleiades cluster \citep{2017A&A...598A..48G} and a centre at $\alpha,\delta =[56.65,24.13]$.  As can be seen from the top panel of this Figure, the DANCe+Tycho catalogue is spatially complete until a radial distance of 11.5 pc ($\sim5^\circ$). The latter corresponds roughly to the sky coverage of the UKIDSS survey \citep{2007MNRAS.379.1599L}. Above this limit, the density of sources drops with two different slopes. The first one is created by the sawtooth pattern at the edge of the DANCe survey, while the last one corresponds to the more extended selection box used for the Tycho survey. To evaluate the photometric completeness, we assume that the distribution of sources in the sky region of the Pleiades is uniform (this simplistic assumption is sufficient for our current purpose). We compare the radial density of sources of different J magnitude bins with that of a synthetic sample uniformly distributed in space and truncated at the completeness radius of 11.5 pc. The radial distribution of this synthetic sample and those of the three magnitude bins are shown in the bottom panel of Fig. \ref{fig:completeness}. As can be seen from the latter, the joint Tycho+DANCe survey is expected to be complete until magnitude $\sim19$ in the J band. Above this limit, the distribution of sources departs significantly from the expected one.  Hence, we restrict our list of candidate members to those with: i) J band observed and less than 19 mag., and ii) radial distances less than 11.5 pc. This results in 1954 candidate members, which represents more than 50\% more candidate members than those of \citet{Converse2010}, who did the latest analysis of the Pleiades PSD.  Accounting for completeness and the previous truncation in the data set is essential to avoid possible bias in the inferred parameters (see Appendix \ref{app:truncation}).  Nevertheless, we remind the reader that the inhomogeneities (e.g. spatial resolutions, gaps in luminosity) of the DANCe+Tycho data set are so complex (and some of them only partially understood) that they can indeed bias the sample of candidate members in unknown ways. For example, the gap in luminosity coverage between the faint end of the Tycho-2 catalogue and the bright end of the DANCe survey (see Fig. 8 of \citealt{2015A&A...577A.148B}) may result in undetected sources, therefore unmeasured proper motions and, finally, an incomplete list of candidate members.

Another important constraint is the number of cluster stars observed within the survey area coverage. 
Truncating the probability distributions properly accounts for the cluster members left outside the truncation radius. However,
due to the several artefacts surrounding the images of bright sources (e.g. halos, spikes, saturation), potential cluster members
could also remain undetected. Furthermore, these artefacts can severely bias any evidence of mass segregation, as the most massive and brightest stars are located at the centre of the cluster. However, the statistical treatment of the impact of these artefacts lays beyond the scope of this work.

The information provided by the observational constraints, which we call $I$, consists of the maximum radius, $R_{max}=11.5$pc, and
the number of stars observed within this radius, $N=1954$. These constraints will be
incorporated into the model through the $likelihood$.

\subsection{Contamination}
\label{sect:contamination}
\cite{Javier} estimate a contamination rate of
$4.3\pm0.2$\% in their sample of candidate members at the probability threshold of $p_{84\%}$\footnote{In \cite{Javier}, individual membership probabilities are themselves probability distributions. Thus, $p_{84\%}$ stands for the 84th percentile of those distributions.}> 0.84. This
would amount to 84 of their 1963 candidate members. Also, \citet{2014A&A...563A..45S} estimate that the contamination rate of their methodology is $11.0\pm2.0$\% for a probability threshold of $p=0.5$, similar to that used by \citet{2015A&A...577A.148B} to classify the candidate members of their Tycho+DANCe data set.  Thus, in our combined Tycho+DANCe list of candidate members, we acknowledge a mean contamination rate of $\sim 8\%$ (approx. 156 objects). We expect these contaminating sources to be uniformly distributed in right ascension and declination because the position on the sky was explicitly removed from the calculation of membership probabilities. Nevertheless, there may be a mild positive gradient of the density towards the Galactic centre. In addition, these contaminants may not be uniformly distributed in $J$ band, with possible concentrations around 14 and 17 mag, where the entanglement of field and cluster populations is higher. The quantification of this possible dependency of contaminants with photometric magnitude and its consequences lay beyond the objective of this work and will be analysed in future studies.

\section{Spatial density models}
\label{sec:models}

\subsection{Spherical models}

In this Section we consider spherically symmetric models of the
spatial distribution of Pleiades members. In the following paragraphs
we give a brief description of each model, its analytical
parameterisation, and the corresponding references.

Our starting point is the classical King's profile. 
Although it was introduced as an empirical law to describe the number surface density of  globular
clusters, it has also been used to describe open clusters
\citep[see][for recent
applications]{2017MNRAS.469.1330A,2017MNRAS.468.2684P}, globular
clusters \citep{2017ApJ...840L..25M} and even to study
galaxies \citep{2017MNRAS.466.1513R}, halo
substructure \citep{2007ApJ...663..960S} and the dark matter
distribution \citep{2016MNRAS.458.2848J}. The analytical description
of the surface number density of stars $n$ is given by

\begin{equation}
  n(R) =
  k\cdot\left(\frac{1}{\sqrt{1+(R/r_c)^2}}-\frac{1}{\sqrt{1+(r_t/r_c)^2}}\right)^2,
\label{eq:King}
\end{equation}

where $r_c$, the core radius, is a scale factor, $r_t$ is the tidal
radius, and $k$ is a constant related (but not equal) to the central
surface density. In the following we use $R$ instead of $r$ (as is
often commonly done in the literature) to refer to the distance from the
system centre projected on the celestial sphere.

In addition to the classical King's profile we have tested two extensions of
it. We define the Generalised King's profile (hereafter GKing) as the
classical King's profile without fixing the exponents of the analytical
expression. Instead of Eq. \ref{eq:King}, we have

\begin{equation}
  n(R)= k \cdot
  \left[
  \left(1+(R/r_c)^{\frac{1}{\alpha}}\right)^{-\alpha}
  -
  \left(1+(r_t/r_c)^{\frac{1}{\alpha}}\right)^{-\alpha}
  \right]^{\beta},
\label{eq:GKing}
\end{equation}

where the classical King's profile is recovered for $\alpha=0.5$ and
$\beta=2$. To the best of our knowledge, only in the work of \citet{2017MNRAS.466.1513R} 
has a similarly modified King's profile been used. However, the profile used by those authors is more restrictive 
than the one presented here, requiring that $\beta = \alpha^{-1}$, and that both terms $(R/r_c)$, and $(r_t/r_c)$ are at the power of 2.

The optimised generalised King's profile (hereafter OGKing) is the GKing
profile with the values of $\alpha$ and $\beta$ fixed at 
the maximum-a-posteriori (MAP) values of the GKing parameters. 
This maximises the Bayesian evidence and reduces the dimensionality
of the parameter space.

To avoid the use of a tidal radius in the radial profile, we have also considered the model proposed by \citet{EFF1987}, henceforth EFF,
to describe young open clusters in the Large
Magellanic Cloud. Their surface density (in star counts per solid
angle) is given by

\begin{equation}
n(R)=n(0)\cdot(1+(R/r_c)^2)^\frac{\gamma}{2},
\label{eq:eff}
\end{equation}
with $r_c$ the core radius, and $\gamma$ the slope of the profile at radii much larger than the core radius.

Finally, we analyse a more general parameterisation introduced in
\cite{1995AJ....110.2622L}, \cite{1996AJ....111.1889B} and
\cite{Zhao1997}, where the projected mass density is given as

\begin{equation}
  \rho(R) = \frac
      {k'}
      {(R/r_c)^{\gamma}\cdot(1+(R/r_c)^{1/\alpha})^{(\gamma-\beta)\alpha}}.
  \label{eq:zhao}
\end{equation}

Equation \ref{eq:zhao} represents a double power law, with $r_c$ being the so-called core or break radius, $\gamma$
and $\beta$ the exponents of the inner and outer regions, respectively, $\alpha$ the width of the transition region, 
and $k'$ a scale constant. 
Meaningful values of these parameters fulfil the following conditions: $\alpha > 0$ and $0\leq \gamma \leq \beta$. 
The aforementioned works assume this functional form for
the projected surface brightness, the projected mass density
$\rho$, and the volume density $v$, although the latter two are
related by integration:

\begin{equation}
  \rho(R) = \int_{0}^{\infty}v(r)\cdot {\rm d}z,  
\end{equation}

where $z$ is the distance along the line of sight.

In this work we use the same analytical expression as in
Eq. \ref{eq:zhao} but for the number density $n(R)$. We call this
model the generalised density profile\footnote{Although it is also
  called Nuker profile by \citet{2010MNRAS.407.2241K}.} (hereafter
GDP), as it comprises many
simpler models, each of which corresponding to particular choices of the model
parameters. Several density profiles proposed to describe galaxies can
indeed be grouped by parameter values. For example,
$\alpha=1$ includes models by \cite{1997ApJ...490..493N},
\cite{1990ApJ...356..359H}, \cite{1983MNRAS.202..995J}, and
\cite{1999MNRAS.310.1147M}. Similarly, $\alpha=1/2, \gamma=0$ includes the
models by \cite{1911MNRAS..71..460P} (with $\beta=5$), by
\cite{1990ApJ...361..408S} and by \cite{1985MNRAS.216..273D}. The EFF
model corresponds also to $\alpha=1/2, \gamma=0$. King's
profile, however, cannot be cast into this general model unless the
tidal radius $r_t$ is fixed at infinity. 

For our spatial analysis, we also considered the restricted
generalised profile (RGDP), corresponding to the generalised profile
with the value $\gamma$ fixed at 0.

We note that we have used similar names for parameters $r_c$ and $\gamma$
in all the aforementioned formulations. However, these parameters do
not share the same meaning amongst models. The latter is distinctively
specified by each model relation $\mathcal{M}$.

In all cases, the $R$ coordinate is defined with respect to the
cluster centre. The actual values of $R$ then depend on
the choice of this origin (see Sect. \ref{sect:centralsym}).

\subsection{Central symmetry constraint}
\label{sect:centralsym}

In the above, we have defined six models: King, GKing,
OGKing, EFF, GDP and RGDP. Each of them has a different set of
parameters. For example, the King's model depends on two parameters
($r_c$ and $r_t$), the EFF model depends on two other parameters ($r_c$
and $\gamma$), and the generalised profile GDP depends on four parameters
($\alpha$, $\beta$, $\gamma$ and $r_c$).

In reality, there are always two more parameters that do not appear
explicitly in any of the above analytical formulations of the
number density profiles. These are the cluster centre
coordinates from which all radial distances $R$ are measured. It is
not a minor question because the problem is degenerate, and there is a
maximum likelihood solution for each choice of the cluster centre. In
principle, one could even choose a poor cluster centre estimate that
renders the angular distribution of members asymmetric, and obtain
a maximum likelihood fit better than those obtained with a better
centred estimate.  The models assume central symmetry, but this can
only be ensured approximately. There is a region of non-negligible
extent, where the cluster centre may be, and any particular choice of
its position will influence the posterior distribution inferred. Thus,
in order to propagate appropriately this uncertainty about the cluster
centre position in our posterior inferences, we have included the two
cluster centre coordinates, $\alpha_c$ and $\delta_c$,
 as further parameters of our models (their allowed intervals will be described in Sect. \ref{sect:priors}).

For any given choice of the central coordinates, 
we calculate the radial distance, $R$, and the position angle $\theta$ 
of each star in our data set. To avoid biases introduced by projection 
effects of objects located far from the cluster centre, we project each 
object's coordinates into the plane of the sky along the line-of-sight 
vector \cite[see for example, Eq. 1 of][]{2006A&A...445..513V}. 

These projected coordinates are
\begin{align}
\tilde{x} =&\sin(\alpha - \alpha_c ) \cdot \cos(\delta),\nonumber\\
\tilde{y} = &\cos(\delta_c)\cdot \sin(\delta) - \sin(\delta_c)\cdot \cos(\delta)\cdot\cos(\alpha - \alpha_c).
\label{eq:distfree}
\end{align}

From these projected coordinates, the radial distance, $R$, and the position angle, $\theta$, are computed as 

\begin{align}
R=& \sqrt{\tilde{x}^2 + \tilde{y}^2},\nonumber\\
\theta=& \arctan2 (\tilde{x},\tilde{y}) +2\pi \ \ (\rm{mod}\ \ 2 \pi).
\label{eq:rs_and_ts}
\end{align}

The requirement of central symmetry is enforced by the inclusion of a
multiplicative term in the likelihood. For a given set of parameter values
of $\alpha_c, \delta_c$ , we divide the computed polar angles of
individual stars, $\theta$, into four symmetric quadrants (divisions at $[0,\pi/2,\pi,3\pi/2]$) and require that the number of
stars in each quadrant be Poisson distributed with a mean rate given
by $N_q=N_{tot}/4$. Under this model, the likelihood of any given
proposal for the model parameters ($\alpha_c$, $\delta_c$)
will be

\begin{align}
\mathcal{L} & = p(N_1,N_2,N_3,N_4|\alpha_c,\delta_c)\nonumber  \\
& =\mathcal{P}(N_1|N_q)\cdot \mathcal{P}(N_2|N_q)\cdot \mathcal{P}(N_3|N_q)\cdot
\mathcal{P}(N_4|N_q), \end{align}

where $N_i, i=1,2,3,4$ is the number of sources in each quadrant, and
$\mathcal{P}(N_i|N_q)$ is the Poisson distribution with mean rate
$N_{tot}/4$ evaluated at $N_i$.

\subsection{Elliptical models}

In this Section we extend the aforementioned spherical models to allow
for deviations from radial symmetry. We do this by allowing variations
of the radial profile that depend on the angular coordinate but still
maintain biaxial symmetry. This can be done in many ways. In this work
we focus on the simplest one: the analytical expression of the radial
profile is maintained along any radial direction but the profile
parameters (e.g. $r_c$ and $r_t$  in the King profile) have an
ellipse-like dependence on the angular coordinate.

This requires the definition of a coordinate system centred at the
cluster centre (parameters $\alpha_c$ and $\delta_c$), and potentially rotated from the RA-Dec system of
axes. Thus, we further include the angle $\phi$ between the principal axes of
the ellipse and RA-Dec system as a parameter of these models. The coordinates $\tilde{x}$ and $\tilde{y}$
of Eq. \ref{eq:distfree} are rotated by angle $\phi$ to obtain coordinates $x$ and $y$. 
Then, $R$ and $\theta$ are computed from the latter by means of Eq. \ref{eq:rs_and_ts}.

The radially symmetric parameters of the previous Section have now an
angular dependency, which is now expressed by means of the characteristic radii at the semi-major and semi-minors axes (denoted by subscripts $a$ and $b$, respectively).
These new radii are expressed as

\begin{equation}
r(\theta) = \frac{r_{a}\cdot r_{b}}{\sqrt{(r_{a}\sin(\theta))^2+r_{b}\cos(\theta))^2}},
\label{eq:rang}
\end{equation}

where $\theta$ is the position angle measured from the semi-major axis, and $r_a$ and $r_b$ are the parameters
representing the characteristic radius at the semi-major and -minor axis, respectively.

We illustrate this new biaxial dependency in the King's profile. 
The surface number density is now
\begin{equation}
  n(R)= k \cdot
  \left(\frac{1}{\sqrt{1+({R}/{r_c(\theta)})^2}} - \frac{1}{\sqrt{1+({r_t(\theta)}/{r_c(\theta)})^2}}\right)^2,
\label{eq:KingEll}
\end{equation}

where $r_c$ and $r_t$ are obtained from Eq. \ref{eq:rang}. Explicitly they are,

\begin{equation}
r_c(\theta) = \frac{r_{ca}\cdot r_{cb}}{\sqrt{(r_{ca}\sin(\theta))^2+r_{cb}\cos(\theta))^2}},
\label{eq:anglerc}
\end{equation}

\begin{equation}
r_t(\theta) = \frac{r_{ta}\cdot r_{tb}}{\sqrt{(r_{ta}\sin(\theta))^2+r_{tb}\cos(\theta))^2,}}
\label{eq:anglert}
\end{equation}

where $r_{ca}$ and $r_{ta}$ are the core and tidal semi-major axis of the ellipse, and $r_{cb}$
and $r_{tb}$ correspond to the semi-minor axis. We highlight that we do not
constrain the two ellipses to have the same aspect ratio, but they are co-aligned.

For the other model, the surface densities are similarly obtained. We
do not incorporate any angle dependence for the exponents $\alpha,\beta$ or $\gamma$.  

The position angle of the semi-major axis with respect to the Right
Ascension axis ($\phi$) is constrained using the equivalent of the
radial symmetry likelihood term, except that now the position angle has
its origin at the semi-major axis.

\subsection{Segregated models}

Finally, in this Section we introduce another set of profiles to
revisit the problem of mass segregation in the
context of the Pleiades.

We consider the previous biaxially symmetric models to which we add
a dependence of the core radius with the J magnitude. We select the J
magnitude because it is the reddest of the magnitudes that are
available for all candidate members. We assume that stars of the same
mass have approximately the same magnitude and that distance
differences (due to the 3D spatial extent of the Pleiades) average
out. The core radius dependence with the J magnitude is modelled as

\begin{equation}
\label{eq:segregation}
  r_c(\theta,J) = r_c(\theta)+\kappa\cdot(J- J_{mode}),
\end{equation}
where $J_{mode}$ is the mode of the J band distribution. 

The slope of the relationship, $\kappa$, is independent of the angle
$\theta$. Therefore, for $J=J_{mode}=13.6$ the
model reduces to the elliptic profile described in Section
\ref{sec:modelselell}. A positive value of $\kappa$ corresponds to
smaller values of the core radius for stars brighter than $J_{mode}=13.6$; in
other words, it describes a system where the more massive stars are
more concentrated than the less massive ones.

\section{Bayesian analysis}
\label{sec:modelselect}

As mentioned in Section \ref{sec:data}, our data set may be contaminated. Thus, in an
effort to minimise the possible impact that these contaminants may
have on our inference, we also model their spatial
distribution. Hence, our model of the spatial distribution of stars
not only includes the model of the Pleiades cluster, but also a field
component which is modelled by a uniform spatial distribution
$\mathcal{U}$ within the maximum radius $R_{max}$.

The measured properties of each of star in our data set can be assumed to be unaffected by the measured properties of any other star in the data set (this assumption is called statistical \textit{independence}). Under this assumption, the probability that the data set was generated by the mixture of cluster and field is the product of the probabilities that each of the stars was generated by this mixture.

Allowing $D=\{d_i,\pi_i\}_i^N$ to denote our data set, with $d$ comprising the sky coordinates and $J$ magnitude,
and $\pi$ the cluster membership probability of each object, the probability or \textit{likelihood} of the data set $D$, given the model $\mathcal{M}$, constraints $I$, and parameters $\bm{q}$, is

\begin{equation}
\mathcal{L}(D|\bm{q},\mathcal{M},I)= \prod_i^N \left[\pi_i\cdot p(d_i|\bm{q},\mathcal{M},I)+(1-\pi_i)\cdot \mathcal{U}(d_i|I) \right].
\end{equation}

The probability $p(d|\bm{q},\mathcal{M},I)$
depends on the profile under consideration and is described in the
following Section.

\subsection{Probabilistic framework}

To avoid the use of bins and to properly infer the parameters of the
models presented in Section \ref{sec:models}, we need to convert the projected stellar densities into
probability density functions that describe the probability of finding
a star between $R$ and $R+{\rm d}R$, under the assumption of spherical
symmetry. The probability density function $p(R)$ is constructed from
the definition:

\begin{equation}
p(R)\cdot {\rm d}R=\frac{2\pi\cdot R \cdot n(R)\cdot {\rm d}R}{N},
\label{eq:probfromdens}
\end{equation}

where $N$ is the total number of stars in the system.

This probability is renormalised to integrate to unity at the truncation radius $R_{max}$, 
which in our data set corresponds to 11.5 pc. Thus,
\begin{equation}
p_T(R)=\left\{
\begin{array}{rcl}
\frac{p(R)}{\int_0^{R_{max}} p(R)\cdot \rm{d}R} &\mbox{for}& R \leq R_{max} \\
0&\mbox{for}& R > R_{max}
\end{array}
\right..
\end{equation}
All the probabilities rendered by our set of models are renormalised according to the previous equation.
However, in the following and for the sake of simplicity, we only present the non-truncated probabilities.

Applying Eq. \ref{eq:probfromdens} to the spherically symmetric
King's profile, we obtain

\begin{equation}
  p(R)=\frac{k\cdot2\pi}{N}\cdot R \cdot
  \left(\frac{1}{\sqrt{1+(R/r_c)^2}} - \frac{1}{\sqrt{1+(r_t/r_c)^2}}\right)^2.
\end{equation}

Actually, in probabilistic inference we write this probability
function as:

\begin{equation}
  p(R|r_c, r_t, k_1,I,\mathcal{M}_1)=k_1\cdot R \cdot
  \left(\frac{1}{\sqrt{1+(R/r_c)^2}} - \frac{1}{\sqrt{1+(r_c/r_t)^2}}\right)^2,
\label{eq:probfromdens1}
\end{equation}

where we have defined a new constant, $k_1=\frac{k\cdot2\pi}{N}$, and
made explicit the dependence of the probability on the underlying
analytical expression ($\mathcal{M}_1$), the constraints $I$, and the
values of the parameter set ($k_1, r_c$ and $r_t$). In practice, $k_1$
is treated as a normalisation constant (to enforce unit integral) and
there is no need to know the total number of stars in the system.

For the generalised King's profile, this becomes
\begin{multline}
  p(R|r_c, r_t, \alpha, \beta, k_2, I, \mathcal{M}_2)= \\
  k_1\cdot R \cdot
  \left[
  \left(1+(R/r_c)^{\frac{1}{\alpha}}\right)^{-\alpha}
  -
  \left(1+(r_t/r_c)^{\frac{1}{\alpha}}\right)^{-\alpha}
  \right]^{\beta}.\\
\end{multline}

Likewise, the expression for the EFF model is

\begin{equation}
  p(R|r_c, \gamma, k_3, I, \mathcal{M}_3)=k_3\cdot R \cdot
  (1+(R/r_c)^2)^\frac{\gamma}{2}.
\label{eq:probfromdens2}
\end{equation}

And finally, the GDP model is given by

\begin{equation}
  p(R| r_c,\alpha, \beta, \gamma, k_4, I, \mathcal{M}_4 ) = \frac
  {k_4\cdot R} {
    (R/r_c)^{\gamma}\cdot(1+(R/r_c)^{1/\alpha})^{(\gamma-\beta)\alpha}},
  \label{eq:probfromdens3}
\end{equation}

with $\gamma=0$ for the RGDP model.

For the elliptical and luminosity segregated density profiles, the
likelihoods are obtained similarly by adding $\phi$ and replacing
$r_c$ and $r_t$ by $r_{ca}, r_{cb}$ and $r_{ta}, r_{tb}$ in the model
parameters, and introducing the dependence on $\theta$ and $J$ in the
relations. For example, the likelihood of the biaxial King's profile
is
\begin{multline}
  p(R,\theta|\phi,r_{ca},r_{cb}, r_{ta},r_{tb}, k_5, I, \mathcal{M}_5)=\\
  k_5\cdot R \cdot
  \left(\frac{1}{\sqrt{1+({R}/{r_c(\theta)})^2}} -
    \frac{1}{\sqrt{1+({r_t(\theta)}/{r_c(\theta)})^2}}\right)^2 \, .
\end{multline}

\subsection{Model selection}

In this Section we aim at comparing the aforementioned analytical
parametrizations of the projected stellar densities in the light of
the currently available data. In order to do so, we use the
Bayesian evidence, also known as marginal likelihood. In the following, we use \emph{evidence} (and its plural \emph{evidences}\footnote{Since the Bayesian evidence is a number that can be computed for each model and/or data set (see Eq. \ref{eq:marlik}), we use the plural \emph{evidences} to address any set containing the Bayesian evidence of more than one model.}) to refer to the Bayesian evidence. The  \emph{evidence} is the key
for model comparison in the Bayesian framework. In this framework, the model
comparison is done on the basis of the model posterior probability
$p(\mathcal{M}|D)$. This is the probability of model $\mathcal{M}$
given the collected data $D$. In our opinion, this is the most natural
way to compare and select (if needed) models in the scientific
context. The posterior probability can be expressed as

\begin{equation}
p({\mathcal M}|D)=\frac{p(D|{\mathcal M})\cdot p({\mathcal M})}{p(D)},
\end{equation}
  
using Bayes' theorem. The ratio of posterior probabilities can then be
expressed as

\begin{equation}
  \frac{p({\mathcal M}_i|D)}{p({\mathcal M}_j|D)}
  =\frac{p(D|{\mathcal M}_i)}{p(D|{\mathcal M}_j)}\cdot
    \frac{p({\mathcal M}_i)}{p({\mathcal M}_j)}.
\end{equation}

If there is no difference in the prior probabilities for models $i$
and $j$, then the posterior ratio is equal to the marginal likelihood
ratio (also known as Bayes Factor), where the marginal likelihood (i.e. the \emph{evidence}) is
the full likelihood marginalised over the model parameters \textbf{q}, as follows

\begin{equation}
p(D|{\mathcal M})=\int p(D|\bf{q},{\mathcal M})\cdot{\rm
  d}\bf{q}.
\label{eq:marlik}
\end{equation}

It is important to remark that the Bayesian model comparison naturally
incorporates a preference towards the less complex models if
they are equally supported by the data. In fact, the preference is
towards models with less effective parameters (understood as
parameters that the data can constrain). 

The computation of the posterior probability distributions and the
\emph{evidence} of each model is carried out in practice using the Nested Sampling
\citep{skilling2006} algorithm as implemented in \emph{PyMultiNest}
\citep{Buchner2014}. 

\subsection{Priors}
\label{sect:priors}
In the spherical models, we have assumed exponential priors, with a scale
value of 1, and truncated at 100, for all exponent parameters $\alpha,
\beta, \gamma$, normal
priors for the central coordinates (with mean at
$[56.65^\circ,24.13^\circ]$ and standard deviation of one degree), and
Half-Cauchy priors for radial parameters (with scale parameter at 10
pc).  These priors fall in the category of \emph{weakly informative}
ones \cite[see][]{Gelman2006}.

In the biaxially symmetric models, we use the same priors as for the
radially symmetric ones but we restrict
the semi-major axes of the core and tidal radii to be larger than, or at
least equal to, their corresponding semi-minor axis. We also include a uniform
prior for the angle $\phi \in [-\pi/2,\pi/2]$.

In the luminosity segregated models, in addition to the previous priors, we use a normal, $\mathcal{N}(0,0.5)$, as a prior for $\kappa$, which represents our prior beliefs of almost negligible luminosity segregation.

The code to perform the analysis of the present work, together with the data set described in Section \ref{sec:data}, is available at \url{https://github.com/olivares-j/PyAspidistra}

\section{Results and Discussion}
\label{sec:results}
We apply the Bayesian formalism described in Section \ref{sec:modelselect} to the data set detailed in Section \ref{sec:data}. Thus, for each of our models we obtain the posterior distribution of its parameters, together with its \emph{evidence}. Appendix \ref{app:posteriors} contains the details of the inferred posterior distributions, figures of the fitted densities and marginal distributions, 
together with the uncertainties of the parameters in each analysed model.
Table  \ref{tab:BFAll} summarises the  \emph{evidences} and Bayes Factors
resulting from all our models and their extensions. In the following we use   these \emph{evidences} to discuss the model comparison.

The boundaries for decision making from
Bayes Factors should be set {\sl ab initio}. We mostly discuss our results following the classical scale
by \cite{Jeffreys61}. In this scale, the strength of the  evidence\footnote{The Jeffreys scale is used to relate the Bayes Factors, which contain the Bayesian evidences of the two models, to the possible shared understanding of the word evidence.} is said to be: {\sl Inconclusive}
if the Bayes Factor is $\lesssim$ 3:1, {\sl weak} if it is $\sim$ 3:1, {\sl moderate} if it is $\sim$12:1, and
{\sl strong} if it is $\gtrsim$ 150 :1. Nevertheless,  we hope that our conclusions can be
shared by the reader independently of the scale used to categorise
the Bayes Factors. 

\begin{table*}
\tabcolsep=1pt
  \caption[]{Natural logarithm of the \emph{evidence} for each
        profile density (diagonal) and Bayes factors (off-diagonal
        elements, with the  \emph{evidence} for the model specified in the
        column header placed in the denominator, i.e. $p(D|\mathcal{M}_{row})/p(D|\mathcal{M}_{column})$). The  \emph{evidence}
        corresponds to the data set truncated at 11.5pc.}  \label{tab:BFAll}
 \resizebox{\textwidth}{!}{
   \begin{tabular}{rr|rrrrrr|rrrrrr|rrrrrr}
\toprule
&{} & \multicolumn{6}{c}{Radial}& \multicolumn{6}{c}{Biaxial}& \multicolumn{6}{c}{Segregated}\\
&{} &      EFF &      GDP &    GKing &     King &   OGKing &     RGDP &      EFF &      GDP &    GKing &     King &   OGKing &     RGDP &      EFF &      GDP &    GKing &     King &   OGKing &     RGDP \\
\midrule
\multirow{6}{*}{\begin{sideways}Radial\end{sideways}}&EFF    & -4569.15 &     8.83 &     0.83 &     0.40 &     0.19 &     2.53 &    <1e-2 &    <1e-2 &    <1e-2 &    <1e-2 &    <1e-2 &    <1e-2 &    <1e-2 &    <1e-2 &    <1e-2 &    <1e-2 &    <1e-2 &    <1e-2 \\
&GDP    &     0.11 & -4571.33 &     0.09 &     0.05 &     0.02 &     0.29 &    <1e-2 &    <1e-2 &    <1e-2 &    <1e-2 &    <1e-2 &    <1e-2 &    <1e-2 &    <1e-2 &    <1e-2 &    <1e-2 &    <1e-2 &    <1e-2 \\
&GKing  &     1.21 &    10.64 & -4568.97 &     0.48 &     0.23 &     3.05 &    <1e-2 &    <1e-2 &    <1e-2 &    <1e-2 &    <1e-2 &    <1e-2 &    <1e-2 &    <1e-2 &    <1e-2 &    <1e-2 &    <1e-2 &    <1e-2 \\
&King   &     2.51 &    22.17 &     2.08 & -4568.23 &     0.49 &     6.35 &    <1e-2 &    <1e-2 &    <1e-2 &    <1e-2 &    <1e-2 &    <1e-2 &    <1e-2 &    <1e-2 &    <1e-2 &    <1e-2 &    <1e-2 &    <1e-2 \\
&OGKing &     5.13 &    45.31 &     4.26 &     2.04 & -4567.52 &    12.99 &    <1e-2 &    <1e-2 &    <1e-2 &    <1e-2 &    <1e-2 &    <1e-2 &    <1e-2 &    <1e-2 &    <1e-2 &    <1e-2 &    <1e-2 &    <1e-2 \\
&RGDP   &     0.40 &     3.49 &     0.33 &     0.16 &     0.08 & -4570.08 &    <1e-2 &    <1e-2 &    <1e-2 &    <1e-2 &    <1e-2 &    <1e-2 &    <1e-2 &    <1e-2 &    <1e-2 &    <1e-2 &    <1e-2 &    <1e-2 \\
\hline
\multirow{6}{*}{\begin{sideways}Biaxial\end{sideways}}&EFF    &     >999 &     >999 &     >999 &     >999 &     >999 &     >999 & -4557.32 &     5.14 &     0.08 &     0.08 &     0.01 &     0.84 &    <1e-2 &    <1e-2 &    <1e-2 &    <1e-2 &    <1e-2 &    <1e-2 \\
&GDP    &     >999 &     >999 &     >999 &     >999 &     >999 &     >999 &     0.19 & -4558.96 &     0.02 &     0.02 &    <1e-2 &     0.16 &    <1e-2 &    <1e-2 &    <1e-2 &    <1e-2 &    <1e-2 &    <1e-2 \\
&GKing  &     >999 &     >999 &     >999 &     >999 &     >999 &     >999 &    12.31 &    63.26 & -4554.81 &     0.97 &     0.13 &    10.37 &    <1e-2 &     0.01 &    <1e-2 &    <1e-2 &    <1e-2 &    <1e-2 \\
&King   &     >999 &     >999 &     >999 &     >999 &     >999 &     >999 &    12.64 &    64.93 &     1.03 & -4554.78 &     0.14 &    10.64 &    <1e-2 &     0.01 &    <1e-2 &    <1e-2 &    <1e-2 &    <1e-2 \\
&OGKing &     >999 &     >999 &     >999 &     >999 &     >999 &     >999 &    91.95 &   472.37 &     7.47 &     7.28 & -4552.80 &    77.41 &     0.04 &     0.10 &    <1e-2 &    <1e-2 &    <1e-2 &     0.04 \\
&RGDP   &     >999 &     >999 &     >999 &     >999 &     >999 &     >999 &     1.19 &     6.10 &     0.10 &     0.09 &     0.01 & -4557.15 &    <1e-2 &    <1e-2 &    <1e-2 &    <1e-2 &    <1e-2 &    <1e-2 \\
\hline
\multirow{6}{*}{\begin{sideways}Segregated\end{sideways}}&EFF    &     >999 &     >999 &     >999 &     >999 &     >999 &     >999 &     >999 &     >999 &   212.43 &   206.96 &    28.45 &     >999 & -4549.45 &     2.95 &     0.10 &     0.03 &     0.16 &     1.15 \\
&GDP    &     >999 &     >999 &     >999 &     >999 &     >999 &     >999 &   886.29 &     >999 &    71.98 &    70.13 &     9.64 &   746.15 &     0.34 & -4550.53 &     0.03 &     0.01 &     0.05 &     0.39 \\
&GKing  &     >999 &     >999 &     >999 &     >999 &     >999 &     >999 &     >999 &     >999 &     >999 &     >999 &   293.70 &     >999 &    10.32 &    30.47 & -4547.12 &     0.32 &     1.64 &    11.86 \\
&King   &     >999 &     >999 &     >999 &     >999 &     >999 &     >999 &     >999 &     >999 &     >999 &     >999 &   913.86 &     >999 &    32.12 &    94.81 &     3.11 & -4545.98 &     5.10 &    36.91 \\
&OGKing &     >999 &     >999 &     >999 &     >999 &     >999 &     >999 &     >999 &     >999 &     >999 &     >999 &   179.23 &     >999 &     6.30 &    18.59 &     0.61 &     0.20 & -4547.61 &     7.24 \\
&RGDP   &     >999 &     >999 &     >999 &     >999 &     >999 &     >999 &     >999 &     >999 &   184.89 &   180.13 &    24.76 &     >999 &     0.87 &     2.57 &     0.08 &     0.03 &     0.14 & -4549.59 \\
\bottomrule
\end{tabular}

 }
  \end{table*}

\subsection{Models with radial symmetry}
\label{sec:modelselrad}
The upper-left panel of Table \ref{tab:BFAll} summarises the  \emph{evidences} and Bayes
Factors obtained from our radially symmetric models. In addition, Table \ref{tab:MAPCtr} shows the Maximum-a-Posteriori (MAP)
 estimate of each parameter in the radially symmetric models
(uncertainties are shown in the Appendix \ref{app:posteriors} in the form of covariance matrices).

 \begin{table}[ht!]
      \caption[]{Maximum-a-posteriori estimates of the inferred parameters in each radially symmetric model.}
         \label{tab:MAPCtr}
         \resizebox{\columnwidth}{!}{
          \begin{tabular}{lrrrrrrr}
\toprule
{} &  $\alpha_c$ [$^\circ$] &  $\delta_c$ [$^\circ$] &  $r_c$ [pc] &  $r_t$ [pc] &  $\alpha$ &  $\beta$ &  $\gamma$ \\
\midrule
EFF    &                  56.66 &                  24.18 &        2.23 &             &           &          &      2.53 \\
GDP    &                  56.66 &                  24.17 &        3.02 &             &      0.64 &     2.95 &      0.09 \\
GKing  &                  56.66 &                  24.16 &        1.42 &       18.17 &      0.46 &     1.48 &           \\
King   &                  56.66 &                  24.16 &        2.04 &       32.08 &           &          &           \\
OGKing &                  56.66 &                  24.17 &        1.38 &       18.87 &           &          &           \\
RGDP   &                  56.66 &                  24.17 &        3.11 &             &      0.69 &     3.13 &           \\
\bottomrule
\end{tabular}

         }
   \end{table}
 
We observe that the  \emph{evidences} cluster in two groups. On one hand there is the family of King's
models, where the  evidence to compare between them is inconclusive and weak in favour of OGKing over GKing. 
On the other hand there are the EFF, GDP, and RGDP, where there is weak  evidence
supporting EFF over GDP and RGDP. There is inconclusive evidence supporting RGDP over GDP.

Comparing the two groups shows that models in King's family provide  evidence that is: inconclusive and weak
over the EFF, weak and moderate over RGDP, and moderate over GDP. Using this information only, we conclude that
the tidal radius is an important parameter.

In addition, we observe that in GDP and RGDP, parameters $r_c$ and $\beta$ show large correlations (0.85 and 0.92 for GDP and RGDP, respectively) and are relatively unconstrained with large uncertainties;  see Appendix \ref{app:posteriors}.
Despite this fact, the models still provide  \emph{evidences} comparable to those of the other models, suggesting that these two parameters, although necessary for the model, are unconstrained by the data, and therefore not penalised by the  \emph{evidence}. Aiming at eliminating this source of degeneracy,
 we tested models in which one of these two parameters was removed. However, the fits and  \emph{evidence} resulting from
them were poorer than that of the RGDP. Thus, we consider these parameters as necessary for this model.

We find that the introduction of more flexibility in the
analytical expressions of the classical radially symmetric profiles
does not provide an increased amount of \emph{evidence}, and results, in some cases, in
unconstrained parameters and a loss of the interpretability associated
to the original formulations. Therefore, the competing models are within the King's family,
with insufficient  evidence to select amongst them. Only
additional, perfectly acceptable prejudices like physical
interpretability or the ability to compare with previous results can
be invoked to choose one (e.g. King's profile) over the rest.

The Bayes Factors seem 
to indicate (with inconclusive evidence however) that the best model is the OGKing.  
However, the fact that this profile has a larger  \emph{evidence} than any of the
remaining models should come as no surprise since it results from fixing
the values of $\alpha$ and $\beta$ of the GKing model to their MAP 
values. 

Comparing the rest of the models, we see that the poorest model is GDP with moderate  evidence against it. 
The best models are again in King's family, followed by EFF and RGDP.

The conclusion from the comparison of these radially symmetric
profiles is that i) there is no
compelling reason to abandon the widely used King profile,
and ii) there are slightly better models, but we lack evidence to prove if they truly 
represent a requirement to make the King's
profile more flexible to accommodate the data. 

\subsection{Biaxially symmetric models}
\label{sec:modelselell}
The central panel of  Table \ref{tab:BFAll} contains the
logarithm of the  \emph{evidences} and Bayes Factors of the biaxially symmetric models. The \emph{evidences} follows a pattern similar to 
that observed for the radially symmetric models, with the exception of those that are against the GDP model. We can conclude that there is
strong evidence for the family of King's models and against the GDP one.
The evidence is still moderate and too weak to compare the rest of the models.

Additionally, we compute a posteriori (from the MCMC chains) the ellipticities\footnote{The ellipticity used here is also known as `flattening'.}  $\epsilon_{rc}$ and $\epsilon_{rt}$, which are defined as,

\begin{align}
\epsilon_{rc} = 1- \frac{r_{cb}}{r_{ca}}, \nonumber \\
\epsilon_{rt} = 1- \frac{r_{tb}}{r_{ta}}, \nonumber
\end{align}
with the latter available only for the King's family of models. 

 \begin{table*}[ht!]
  \centering
      \caption[]{Maximum-a-posteriori estimates of the inferred parameters in each biaxially symmetric model. Ellipticities are derived a posteriori using the inferred parameters.}
         \label{tab:MAPEll}
          \begin{tabular}{lrrrrrrrrrrrr}
\toprule
{} &  $\alpha_c$ [$^\circ$] &  $\delta_c$ [$^\circ$] &  $\phi$ [rad] &  $r_{ca}$ [pc] &  $r_{ta}$ [pc] &  $r_{cb}$ [pc] &  $r_{tb}$ [pc] &  $\alpha$ &  $\beta$ &  $\gamma$ &  $\epsilon_{rc}$ &  $\epsilon_{rt}$ \\
\midrule
EFF    &                  56.66 &                  24.15 &          0.99 &           2.61 &                &           2.11 &                &           &          &      2.58 &             0.22 &                  \\
GDP    &                  56.64 &                  24.15 &          1.01 &           3.90 &                &           3.14 &                &      0.68 &     3.28 &      0.04 &             0.23 &                  \\
GKing  &                  56.66 &                  24.14 &          0.94 &           1.35 &          18.00 &           1.21 &          12.79 &      0.48 &     1.34 &           &             0.10 &             0.30 \\
King   &                  56.64 &                  24.20 &          1.01 &           2.05 &          51.23 &           2.04 &          20.92 &           &          &           &             0.07 &             0.64 \\
OGKing &                  56.68 &                  24.16 &          1.04 &           1.51 &          22.63 &           1.38 &          14.54 &           &          &           &             0.09 &             0.36 \\
RGDP   &                  56.68 &                  24.17 &          0.96 &           4.05 &                &           3.04 &                &      0.78 &     3.32 &           &             0.24 &                  \\
\bottomrule
\end{tabular}

   \end{table*}
   
Table \ref{tab:MAPEll} shows the MAP estimate for the parameters in the models of this Section, together with the mode of the distributions of ellipticities. Uncertainties for the latter are given in Appendix \ref{app:posteriors}.

We can observe that models with no tidal radius have similar $\epsilon_{rc}$ ellipticities with a mean value of $0.23\pm0.01$. This value is similar to the 0.17  found by \citet{Raboud1998}, who use a multicomponent analysis to derive the directions (although its value is not given) and the aspect ratio of the ellipse's axes. However, it is very interesting to see that the models within King's family result in lower values of the ellipticity in the central region and larger values in the outer one. This result is expected from the interaction with the galactic potential and is predicted by numerical simulations of open clusters \cite[see, e.g.][]{1987MNRAS.224..193T}.

By comparing the \emph{evidences} of the biaxially symmetric models to those of the radially symmetric ones, we can conclude that in all cases there is strong  evidence in favour of the biaxial models.

\subsection{Models with luminosity segregation}
\label{sec:lumin-segreg}
The lower-right panel of Table \ref{tab:BFAll} summarises the  \emph{evidences} and Bayes
Factors of models with luminosity segregation. Also, Table \ref{tab:MAPSg} shows the MAP of the inferred distributions for this set of models, together with the derived ellipticities.

\begin{table*}[ht!]
  \centering
      \caption[]{Maximum-a-posteriori estimates of the inferred parameters in each luminosity segregated model. Ellipticities are derived a posteriori using the inferred parameters.}
         \label{tab:MAPSg}
          \begin{tabular}{lrrrrrrrrrrrrr}
\toprule
{} &  $\alpha_c$ [$^\circ$] &  $\delta_c$ [$^\circ$] &  $\phi$ [rad] &  $r_{ca}$ [pc] &  $r_{ta}$ [pc] &  $r_{cb}$ [pc] &  $r_{tb}$ [pc] &  $\alpha$ &  $\beta$ &  $\gamma$ &  $\kappa$ [pc mag$^{-1}$] &  $\epsilon_{rc}$ &  $\epsilon_{rt}$ \\
\midrule
EFF    &                  56.66 &                  24.16 &          1.02 &           2.65 &                &           2.22 &                &           &          &      2.60 &                      0.12 &             0.18 &                  \\
GDP    &                  56.68 &                  24.17 &          1.01 &           3.60 &                &           3.19 &                &      0.63 &     3.14 &      0.13 &                      0.23 &             0.18 &                  \\
GKing  &                  56.66 &                  24.16 &          0.83 &           1.39 &          16.88 &           1.22 &          12.61 &      0.67 &     1.28 &           &                      0.13 &             0.05 &             0.38 \\
King   &                  56.62 &                  24.19 &          0.96 &           2.34 &          38.49 &           2.37 &          20.49 &           &          &           &                      0.19 &             0.05 &             0.60 \\
OGKing &                  56.61 &                  24.17 &          0.99 &           1.62 &          22.08 &           1.59 &          14.04 &           &          &           &                      0.10 &             0.07 &             0.36 \\
RGDP   &                  56.62 &                  24.17 &          0.96 &           3.78 &                &           3.35 &                &      0.73 &     3.34 &           &                      0.24 &             0.19 &                  \\
\bottomrule
\end{tabular}

   \end{table*}
   
 We observe that the ellipticities follow the same pattern as those of the previous Section. This is expected because we explicitly model the luminosity segregation as independent of the position angle.
   
The luminosity segregation inferred here is non negligible with $\kappa$ in the range 0.1 to 0.25 $\rm{pc}\,\rm{mag}^{-1}$, thus indicating that it is indeed an important parameter. However, in all the models, the marginal posterior distribution of $\kappa$ does not discard the zero value (see the marginal posterior of $\kappa$ in Appendix \ref{app:posteriors}).

The \emph{evidences} provided by the models with luminosity segregation follow a similar pattern as those from radial symmetry. However, in this case the best model is the classical King's, which shows only moderate evidence against the EFF, RGDP, and GDP models. The evidence of King's model over GKing and OGKing is weak.

The \emph{evidences} provided by the luminosity segregated models lead to them being strongly favoured over the radially and biaxially symmetric ones in all cases. We can conclude that, despite having a small value of $\kappa,$ the luminosity segregation is an important parameter regardless of the model used.

\subsection{Total mass and number of members }
\label{sec:mass}

In this Section we use the inferred values of the parameters in King's family of models to derive simple estimates of the total number of members and mass
of the Pleiades cluster.

\begin{figure}[ht!]
\begin{center}
  \includegraphics[width=\columnwidth]{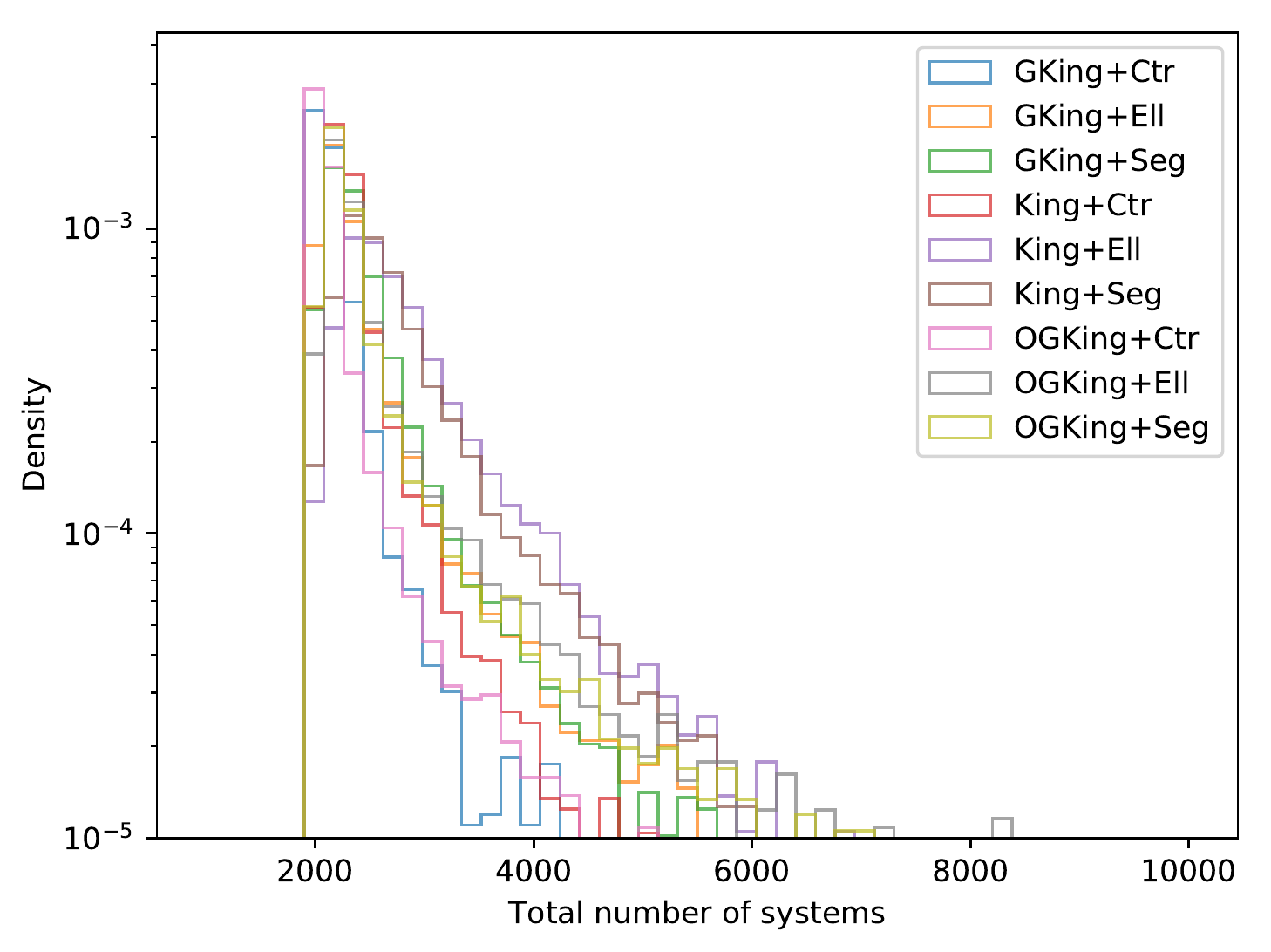}
\caption{Distribution of the total number of systems within the tidal radius in each model and extension of the King's family.
The abbreviations Ctr, Ell, and Seg stand for the radial and biaxial symmetric models, and those with luminosity segregation, respectively.}
\label{fig:dist_numbers}
\end{center}
\end{figure}

For each model and extension within the King's family, we estimate the total number of cluster members by integrating the surface density profile until the tidal radii inferred for the model. This is done for each set of parameters returned by \emph{PyMultiNest}. The resulting distributions of the total numbers fore each model and extension in the King's familly are shown in Fig. \ref{fig:dist_numbers}. Additionally, Table \ref{tab:numbers} shows the mode of these distributions. As can be seen from this Table, our current data set (with 1954 members), although twice as large as previous studies in the literature, still lacks almost one fifth of the predicted number of objects in the cluster.

\begin{table}[ht!]
\centering
\caption{Mode of the distribution of total number of stars in the cluster.}
 \begin{tabular}{lrrr}
\toprule
{} &  GKing &  King &  OGKing \\
\midrule
Ctr &   2087 &  2251 &    2086 \\
Ell &   2209 &  2509 &    2257 \\
Seg &   2272 &  2455 &    2231 \\
\bottomrule
\end{tabular}

\label{tab:numbers}
\end{table}%

We also estimated the total mass of the cluster using the posterior samples of the parameters returned by \emph{PyMultiNest}. To gain an estimate of the total mass we use the tidal force resulting from the interaction of the self-gravitating cluster with the galactic potential. A detailed derivation of the Jacobi radius under the Hill's approximation can be found at  p. 681 of \citet{2008gady.book.....B}. Following the mentioned authors, the Jacobi radius is given by,

\begin{equation}
\label{eq:jacobi}
r_J = \left(\frac{Gm}{4\Omega_0A_0}\right)^{1/3},
\end{equation}
where $G$ is the gravitational constant, $m$ the total mass of the cluster, and $\Omega_0$ the circular frequency of the cluster around the galactic centre, which can be expressed in terms of the Oort's constants $A_0$ and $B_0$ as  $\Omega_0=A_0 - B_0$.  

In the following, we assume an over-simplistic correspondence between the tidal radius of the King's family of models and Jacobi radius.  \citet[p. 677]{2008gady.book.....B} provide a detailed list of reasons why this correspondence is only approximate. Thus, using the Oort's constant values given by \citet[$A=15.3\pm0.4 \,\rm{km s}^{-1} \rm{kpc}^{-1}$ and $B=-11.9\pm0.4 \,\rm{km s}^{-1} \rm{kpc}^{-1}$]{2017MNRAS.468L..63B}, we can derive an estimate of the total mass of the cluster for each inferred value of the tidal radius. 

\begin{figure}[ht!]
\begin{center}
  \includegraphics[page=3,width=\columnwidth]{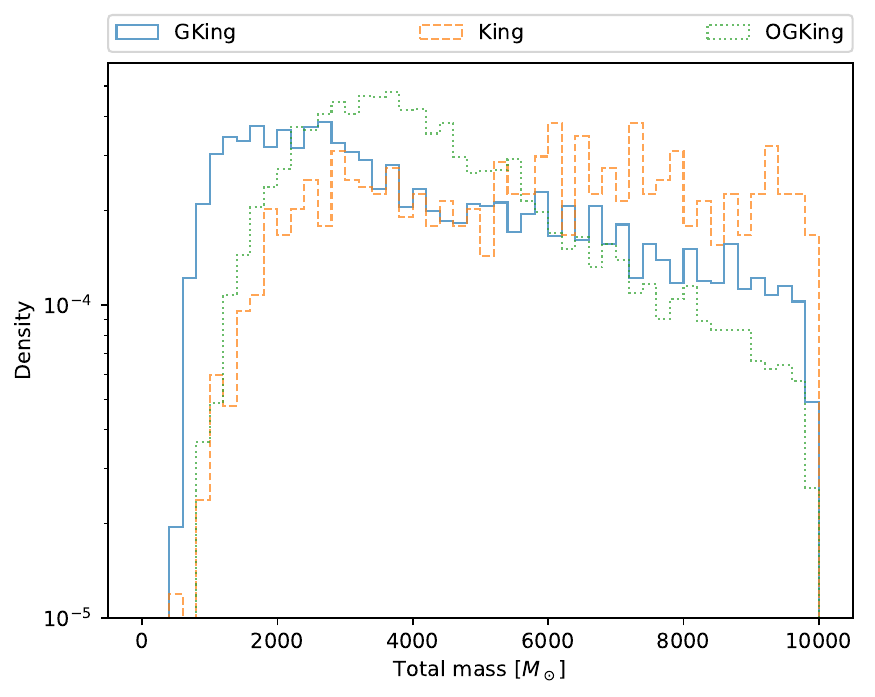}
\caption{Distribution of the total mass of the cluster derived from each biaxially symmetric and luminosity segregated model of the King's family.}
\label{fig:mass}
\end{center}
\end{figure}

Figure \ref{fig:mass} shows the distributions of the total mass derived from the posterior distributions of the parameters of the King's family of models with biaxial symmetry and luminosity segregation (the distributions of total mass resulting from the radial and biaxial models are shown in Appendix \ref{app:posteriors}). As a summary, Table \ref{tab:masses} shows the mode of each of these total mass distributions. 

As can be seen from this Figure and Table, inferring the total mass by means of the poorly constrained tidal radius leads to large uncertainties and probably biased estimators. This effect has already been observed by \citet{Raboud1998}, who derived a total mass of  4000 M$_\odot$ with a confidence interval  ranging from 1600 M$_\odot$ to 8000 M$_\odot$. These values are in good agreement with the ones reported in Table \ref{tab:masses} and observed in Fig. \ref{fig:mass}. 

Given the large ellipticity of the cluster, we also investigated the possibility of deriving the total mass by means of the tidal elongation effect.
However, the values determined are even more poorly constrained than those determined using Eq. \ref{eq:jacobi}. 

The results of this Section show that: i) there is still a large fraction (up to 20\%) of cluster members that lay beyond the spatial coverage of our data set, and, ii) although poorly unconstrained, the distributions of the total mass of the cluster seem to suggest that it is highly unlikely that the total mass of the cluster lays below the 1000  M$_\odot$ limit, as commonly stated in the literature. However, the large and unconstrained mass distribution could also be an artefact  resulting from: i) the poor correspondence between the Jacobi radius and the tidal radius, ii) the poorly constrained values of the tidal radius, and iii) dynamical effects not taken into account to derive Eq. \ref{eq:jacobi} (e.g. the cluster is not a point mass but a self gravitating and rotating system).

\begin{table}[ht!]
\centering
\caption{Mode of the distribution of total mass of the cluster. Units in solar masses.}
 \begin{tabular}{lrrr}
\toprule
{} &  GKing &  King &  OGKing \\
\midrule
Ctr &   1408 &  8584 &    2277 \\
Ell &   1956 &  6049 &    3571 \\
Seg &   2247 &  6605 &    3508 \\
\bottomrule
\end{tabular}

\label{tab:masses}
\end{table}%

\section{Conclusions}
\label{sec:conclusions}
In this work we have formulated the existing radially symmetric
alternatives for the spatial distribution of stars in open clusters in
a probabilistic framework. The set of distributions reviewed include
i) the classical King's profile with two variants put forward by us, ii) the EFF
model, and, iii)  a general profile inspired by
galactic profiles together with a more restricted version of it.  We
have used Bayesian techniques to both obtain posterior probability
distributions for the parameters, and \emph{evidences} for each model. With them, we compare and
select the best model, given the data (and its possible biases). Furthermore, we have computed 
Bayes Factors for all pairwise model comparisons. Due to high correlations among
their $r_c$ and $\beta$ parameters, the GDP and RGDP models loose their physical interpretability. 
The result of the comparison amongst models with radial symmetry is that the King's family of models is only 
mildly superior, with weak and moderate evidence, to those models without the tidal radius parameter.

Furthermore, we have analysed biaxially symmetric extensions of our set of models. 
The results indicate that deviations from spherical symmetry have strong  evidence when compared to the
more simple radially symmetric models. Additionally, the distribution of ellipticities derived from the EFF, GDP, and RGDP
models peak at $0.22\pm0.01$, which is similar to the value of 0.17  found by \citep{Raboud1998}.
Within the King's family, the models return ellipticities that are small (mean $\epsilon_{rc}= 0.07\pm0.02$) and large (mean $\epsilon_{rt}=0.44\pm0.14$) in the inner and outer parts of the cluster, respectively. This effect is expected from the dynamical interaction of the cluster with the galactic potential, and is also predicted by numerical simulations. 

We use Bayesian model selection with Bayes Factors to analyse
mass segregation. We prefer to remain
in the domain of direct observables and study potential differences in
the parameters of the spatial distribution as a function not of mass,
but of the apparent J-band magnitude. The Bayes Factors show strong evidence in favour of the luminosity segregated models, and against the simpler biaxially symmetric ones. We interpret this result as strong evidence for mass segregation. 

The above conclusions heavily depend on the sample of Pleiades members
selected for the analysis. In our probabilistic analysis we took
into account the possibility that our sample is contaminated, but a
$J$-band-dependent contamination rate ($J$-band contamination gradient) could mimic a mass segregation
such as the one observed here. In addition, the halos and artefacts in the images of the central and bright stars
can induce a spatial incompleteness that could also artificially enhance the slope of the luminosity segregation.
Thus, our results must be taken with care. In the near future, we expect to conduct similar studies given the 
more homogenous and well characterised data sets (e.g. new releases of \emph{Gaia}'s data).

Although the the GKing and OGKing models introduced here
have greater \emph{evidences} and fitting properties than the classical King's profile, there is no strong evidence supporting an abandonment of the latter. 
Nevertheless, the GKing profile is a good alternative to the King's classical profile and should be compared with it
 in light of new and more complete data sets.

From the model selection process, we can conclude that the classical King's profile extended to include biaxial symmetry and mass/luminosity segregation
should be the starting point in future analyses of the spatial distribution of open clusters.

Finally, we use the posterior distributions of the parameters in King's model family to obtain rough estimates of the total mass and number of systems in the cluster. We observe that even the largest census of candidate members \citep{2015A&A...577A.148B,Javier} may lack up to 20\% of the predicted number of stellar systems. The probability distribution function of the cluster total mass, which is determined using approximations of the tidal force exerted by the galactic and cluster potentials, reveals that it is highly unlikely that the true cluster total mass lays below the 1000 M$_\odot$ limit. 

The results of this work suggest that, although the Pleiades 
cluster is one of the most studied in the literature, the daughters of Atlas still keep
many of their secrets within the oceans of the sky; probably awaiting the arrival of the final \emph{Gaia}'s data.


\begin{acknowledgements}
We express our gratitude to the referee Anthony G.A. Brown for his kind and assertive comments which greatly improved the quality of this work. 

This research has received funding from the European Research Council (ERC) under the European Union’s Horizon 2020 research 
and innovation programme (grant agreement No 682903, P.I. H. Bouy), and from the French State in the framework of the 
"Investments for the future" Program, IdEx Bordeaux, reference ANR-10-IDEX-03-02. D. Barrado acknowledges the support by the Spanish project  ESP2015-65712-C5-1-R. E. Moraux acknowledges financial support from  the "StarFormMapper" project funded by the European Union’s Horizon 2020 Research and Innovation Action (RIA) programme under grant agreement number 687528. This work has been partially supported by a cooperative PICS project between CNRS and CSIC.
\end{acknowledgements}

\bibliographystyle{aa} 
\bibliography{olivares} 

\begin{appendix}
\section{Effects of truncation on King's profile.}
 \label{app:truncation}
 
Statistical truncation occurs when an unknown number of sources lay beyond a threshold value. This threshold value can originate in the measuring process or in the post-processing of the data. The resulting data set does not contain any information about objects beyond the threshold.

\begin {figure}[ht!]
 \centering
    \includegraphics[page=1,width=\columnwidth]{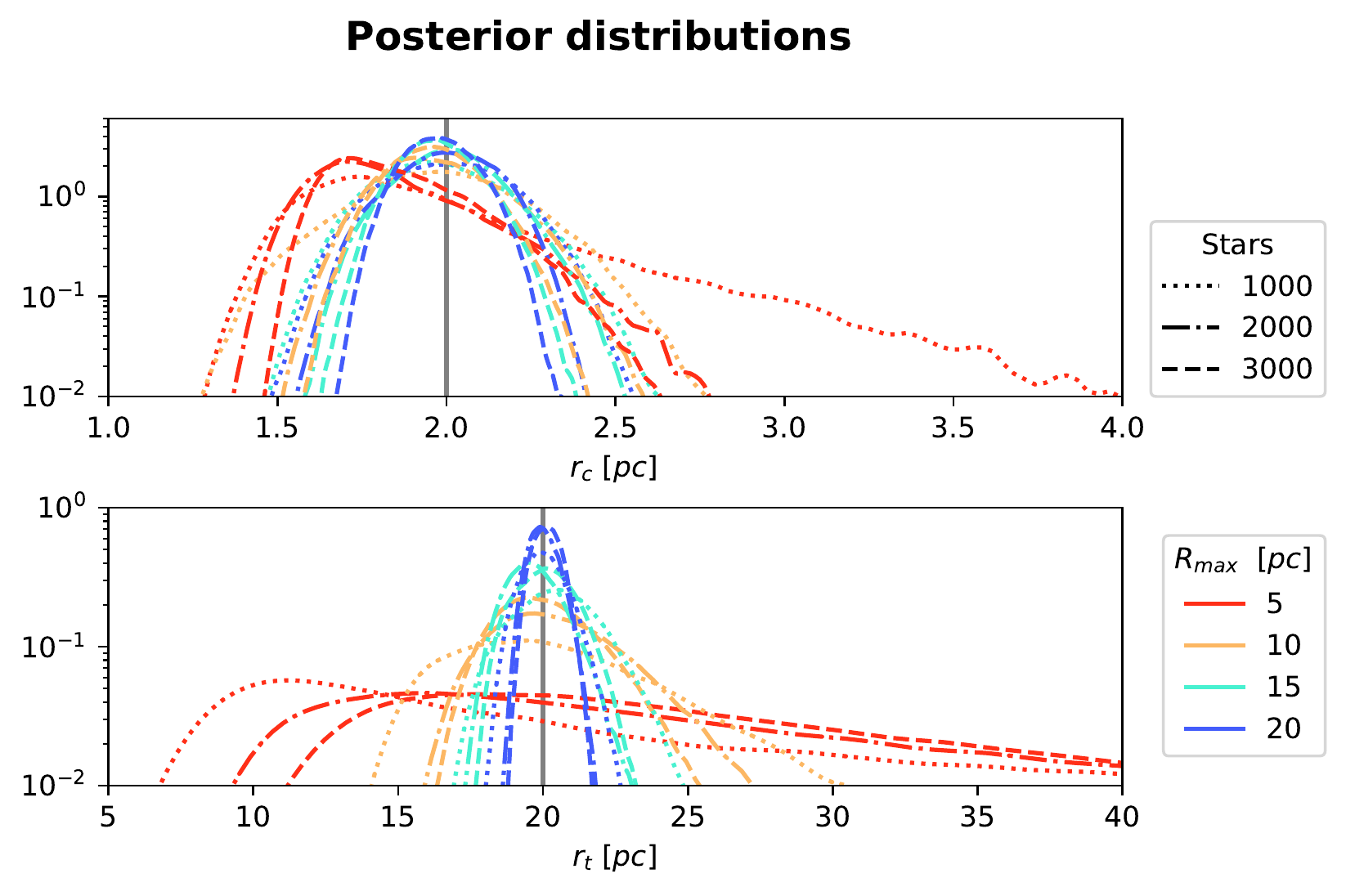}
  \caption{Mixture of the ten posterior distributions of the core and tidal radius ($r_c$ and $r_t$, respectively) inferred under different sample sizes (line styles) and truncation radii (colours). The true parameter values are shown with the vertical grey lines.}
\label{fig:KingSyn}
\end {figure}

\begin {figure}[ht!]
 \centering
    \includegraphics[page=2,width=\columnwidth]{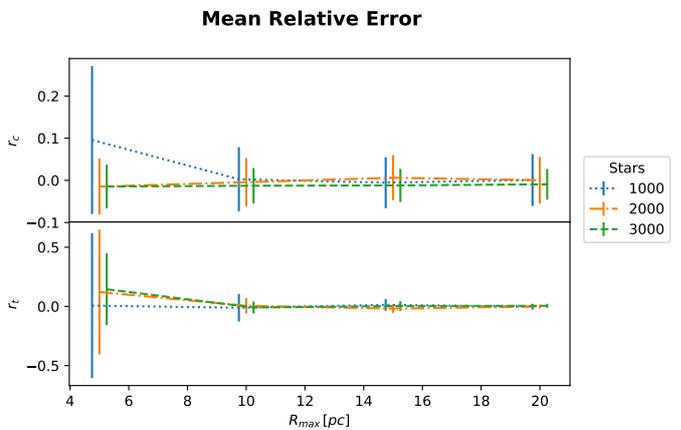}
  \caption{Mean relative error ($r_c$ and $r_t$, respectively) of the MAP statistic inferred from ten random realisations of different sample sizes (line styles) and truncation radii (colours). The uncertainties correspond to the standard deviation of the ten inferred MAPs.}
\label{fig:KingSynMRE}
\end {figure}

Performing inference on truncated data can bias the recovered parameters if the truncation mechanism is not included in the analysis. Nevertheless, bias can still appear if poor statistics are used to summarise the results. Practically speaking, if the truncation is too restrictive it could also lead to bias due to a reduced sample size. To estimate the impact of these effects, we generated synthetic data sets from the King's profile, at true values of $r_c =2.0$ pc and $r_t=20.0$ pc, and infer the parameters under different sample sizes (1000,2000, and 3000 objects) and truncation radii (5,10,15,20 pc). We repeat each estimation ten times to account for randomness in the sample. Figure \ref{fig:KingSyn} shows the posterior distributions inferred at each sample size and truncation radius. As can be seen, accounting for truncation results in posterior distribution that correctly recovers the true parameter values. However, due to the large asymmetry in the posterior distributions of the tidal radius at the lower truncation radius (5 pc), the Maximum A Posteriori (MAP) statistic can be severely biased. Figure \ref{fig:KingSynMRE} shows the mean relative error of this statistic as a function of the truncation radius. As can be seen, the larger biases appear at the extreme case where the truncation radius is only one fourth of the true tidal radius. We note that although the MAP estimates of each of the ten realisations are biased, estimates are made in a similar way above and below the true value; except at the truncation radius of 5 pc, where they slightly over estimate the value. Also, the MAP is unbiased above truncation radii of half the tidal radius, in spite of the number of stars (at least for the tested values).

This example shows that the inference of the parameters in the King's profile can be biased even after truncation has been accounted for. In particular, the tidal radius can be severely affected by truncation radius below one half of the tidal radius. Since this phenomenon is observed under the weakly informative priors used (half-cauchy centred at zero and scale parameter of 100), this effect can be generalised to any maximum-likelihood estimator, the $\chi^2$ statistic particularly. 

\begin {figure}[ht!]
 \centering
    \includegraphics[page=1,width=\columnwidth]{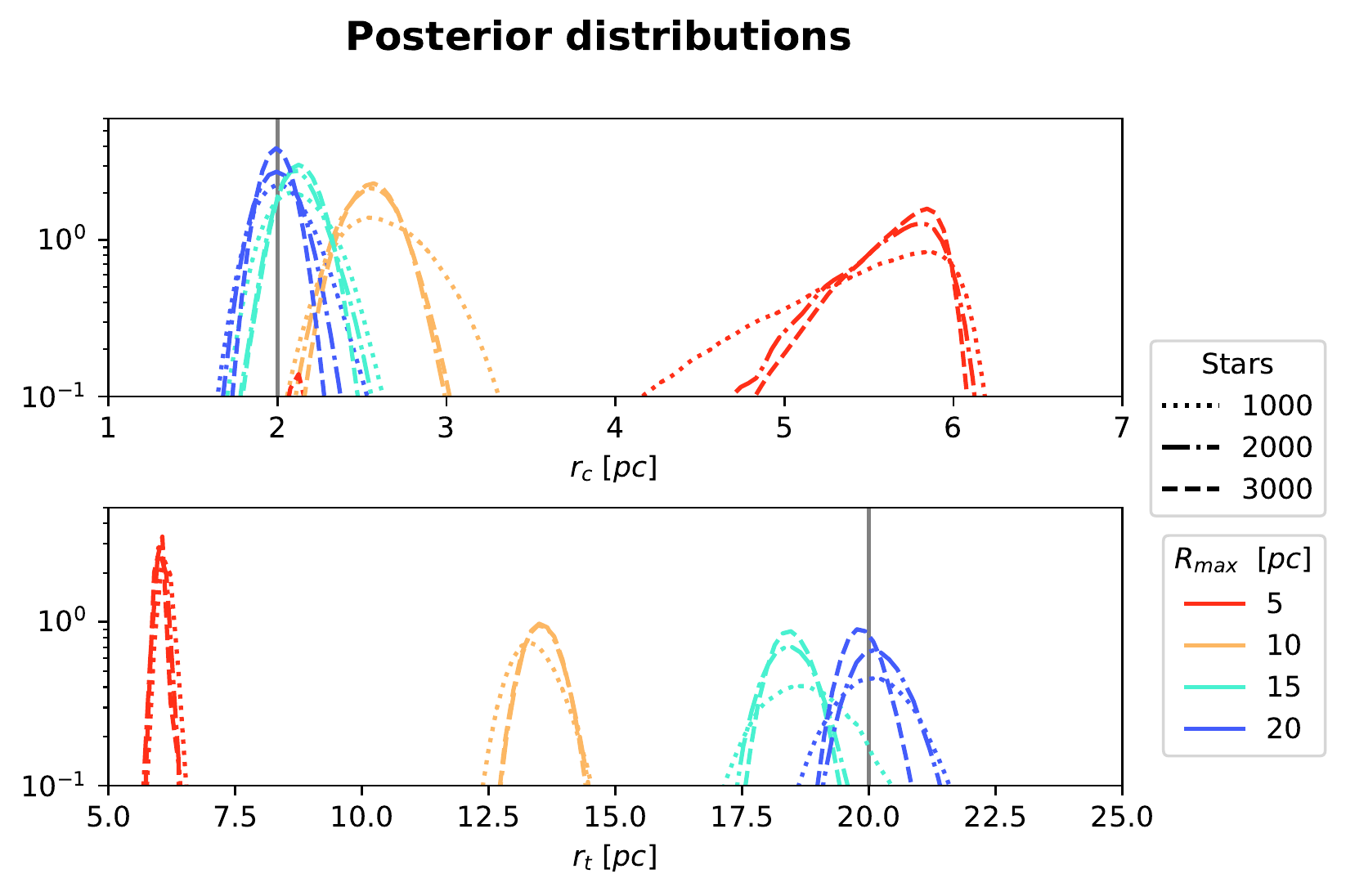}
  \caption{Mixture of the ten posterior distributions of the core and tidal radius ($r_c$ and $r_t$, respectively) inferred under different sample sizes (line styles) and truncation radii (colours) without correcting for truncation. The true parameter values are shown with the vertical grey lines.}
\label{fig:KingSynNoTrunc}
\end {figure}

Finally, as can be seen in Fig. \ref{fig:KingSynNoTrunc}, inferring King's profile parameters without properly accounting for truncation leads to even larger biases.

\section{Posterior distributions}
 \label{app:posteriors}
 
 This Appendix contains the details of the inference performed for each of the models and extensions presented in Section \ref{sec:models}.
 It is structured in the same way as that Section. It starts with the radial models, then continues with the biaxial extensions, and finishes with the
 luminosity segregated ones. For each extension we give: i) the uncertainties of the MAP for each model, and, ii) Figures depicting:  a) the number surface density (i.e. the number of stars per square parsec), and, b) the univariate and bivariate marginal posterior distributions obtained from \emph{PyMultiNest} in the form of a corner plot \citep{corner}. Since the MAP is computed in the joint posterior, it does not necessarily coincides with the modes of the
marginal distributions. 

The MAP uncertainties and correlations are summarised by covariance matrices. These
are computed using the 68.2\% of samples from the MCMC that were the closest to the MAP value. They represent the $2\sigma$ uncertainties
and correlations of the parameters at the vicinity of the MAP. 

For the biaxial and luminosity segregated models we also give the ellipticity distributions computed a posteriori from the core and tidal (when available) semi-major and semi-minor axes resulting from the \emph{PyMultiNest} samples.

Finally, this Appendix also contains the distributions of the total mass of the cluster derived from the radial and biaxial models in the King's family.

 \begin{table}
      \caption[]{Covariance matrix of the radially symmetric EFF model.}
         \resizebox{\columnwidth}{!}{
           \begin{tabular}{lrrrr}
\toprule
{} &  $\alpha_c\ \ [^\circ]$ &  $\delta_c\ \ [^\circ]$ &  $r_c$ [pc] &  $\gamma$ \\
\midrule
$\alpha_c\ \ [^\circ]$ &                   0.007 &                   0.000 &       0.000 &     0.000 \\
$\delta_c\ \ [^\circ]$ &                   0.000 &                   0.006 &       0.001 &     0.000 \\
$r_c$ [pc]             &                   0.000 &                   0.001 &       0.058 &     0.030 \\
$\gamma$               &                   0.000 &                   0.000 &       0.030 &     0.027 \\
\bottomrule
\end{tabular}

         }
   \end{table}
 
 \begin{table}
      \caption[]{Covariance matrix of the radially symmetric GDP model.}
         \resizebox{\columnwidth}{!}{
  \begin{tabular}{lrrrrrr}
\toprule
{} &  $\alpha_c\ \ [^\circ]$ &  $\delta_c\ \ [^\circ]$ &  $r_c$ [pc] &  $\alpha$ &  $\beta$ &  $\gamma$ \\
\midrule
$\alpha_c\ \ [^\circ]$ &                   0.012 &                   0.000 &       0.004 &     0.000 &    0.002 &     0.000 \\
$\delta_c\ \ [^\circ]$ &                   0.000 &                   0.012 &       0.006 &     0.002 &    0.006 &    -0.001 \\
$r_c$ [pc]             &                   0.004 &                   0.006 &       0.589 &     0.062 &    0.330 &     0.004 \\
$\alpha$               &                   0.000 &                   0.002 &       0.062 &     0.046 &    0.059 &    -0.016 \\
$\beta$                &                   0.002 &                   0.006 &       0.330 &     0.059 &    0.256 &    -0.028 \\
$\gamma$               &                   0.000 &                  -0.001 &       0.004 &    -0.016 &   -0.028 &     0.028 \\
\bottomrule
\end{tabular}

         }
   \end{table}

 \begin{table}
  \centering
      \caption[]{Covariance matrix of the radially symmetric GKing model.}
         \resizebox{\columnwidth}{!}{
  \begin{tabular}{lrrrrrr}
\toprule
{} &  $\alpha_c\ \ [^\circ]$ &  $\delta_c\ \ [^\circ]$ &  $r_c$ [pc] &  $r_t$ [pc] &  $\alpha$ &  $\beta$ \\
\midrule
$\alpha_c\ \ [^\circ]$ &                   0.022 &                   0.001 &       0.002 &      -0.021 &     0.001 &   -0.000 \\
$\delta_c\ \ [^\circ]$ &                   0.001 &                   0.019 &       0.007 &       0.046 &    -0.004 &    0.005 \\
$r_c$ [pc]             &                   0.002 &                   0.007 &       1.376 &       1.469 &     0.126 &    0.317 \\
$r_t$ [pc]             &                  -0.021 &                   0.046 &       1.469 &      19.684 &    -0.214 &    1.139 \\
$\alpha$               &                   0.001 &                  -0.004 &       0.126 &      -0.214 &     0.364 &    0.027 \\
$\beta$                &                  -0.000 &                   0.005 &       0.317 &       1.139 &     0.027 &    0.141 \\
\bottomrule
\end{tabular}

         }
   \end{table}

 \begin{table}
      \caption[]{Covariance matrix of the radially symmetric King model.}
         \resizebox{\columnwidth}{!}{
  \begin{tabular}{lrrrr}
\toprule
{} &  $\alpha_c\ \ [^\circ]$ &  $\delta_c\ \ [^\circ]$ &  $r_c$ [pc] &  $r_t$ [pc] \\
\midrule
$\alpha_c\ \ [^\circ]$ &                   0.018 &                   0.001 &       0.001 &      -0.001 \\
$\delta_c\ \ [^\circ]$ &                   0.001 &                   0.017 &       0.002 &       0.021 \\
$r_c$ [pc]             &                   0.001 &                   0.002 &       0.144 &      -0.945 \\
$r_t$ [pc]             &                  -0.001 &                   0.021 &      -0.945 &      37.437 \\
\bottomrule
\end{tabular}

         }
   \end{table}

 \begin{table}
      \caption[]{Covariance matrix of the radially symmetric OGKing model.}
         \resizebox{\columnwidth}{!}{
  \begin{tabular}{lrrrr}
\toprule
{} &  $\alpha_c\ \ [^\circ]$ &  $\delta_c\ \ [^\circ]$ &  $r_c$ [pc] &  $r_t$ [pc] \\
\midrule
$\alpha_c\ \ [^\circ]$ &                   0.011 &                   0.001 &      -0.001 &      -0.003 \\
$\delta_c\ \ [^\circ]$ &                   0.001 &                   0.010 &      -0.000 &      -0.000 \\
$r_c$ [pc]             &                  -0.001 &                  -0.000 &       0.054 &      -0.100 \\
$r_t$ [pc]             &                  -0.003 &                  -0.000 &      -0.100 &       1.951 \\
\bottomrule
\end{tabular}

         }
   \end{table}

 \begin{table}
      \caption[]{Covariance matrix of the radially symmetric RGDP model.}
         \resizebox{\columnwidth}{!}{
  \begin{tabular}{lrrrrr}
\toprule
{} &  $\alpha_c\ \ [^\circ]$ &  $\delta_c\ \ [^\circ]$ &  $r_c$ [pc] &  $\alpha$ &  $\beta$ \\
\midrule
$\alpha_c\ \ [^\circ]$ &                   0.014 &                   0.000 &       0.003 &    -0.000 &    0.001 \\
$\delta_c\ \ [^\circ]$ &                   0.000 &                   0.012 &       0.006 &     0.000 &    0.005 \\
$r_c$ [pc]             &                   0.003 &                   0.006 &       0.804 &     0.089 &    0.466 \\
$\alpha$               &                  -0.000 &                   0.000 &       0.089 &     0.051 &    0.061 \\
$\beta$                &                   0.001 &                   0.005 &       0.466 &     0.061 &    0.311 \\
\bottomrule
\end{tabular}

         }
   \end{table}
   
\begin{figure*}
\subfloat[][EFF]{   \includegraphics[page=2,width=0.5\textwidth]{Analysis/Centre/EFF_11/Ctr_EFF_fit.pdf}}
\subfloat[][GDP]{   \includegraphics[page=2,width=0.5\textwidth]{Analysis/Centre/GDP_11/Ctr_GDP_fit.pdf}}
\\
\subfloat[][GKing]{ \includegraphics[page=2,width=0.5\textwidth]{Analysis/Centre/GKing_11/Ctr_GKing_fit.pdf}}
\subfloat[][King]{  \includegraphics[page=2,width=0.5\textwidth]{Analysis/Centre/King_11/Ctr_King_fit.pdf}}
\\
\subfloat[][OGKing]{\includegraphics[page=2,width=0.5\textwidth]{Analysis/Centre/OGKing_11/Ctr_OGKing_fit.pdf}}
\subfloat[][RGDP]{  \includegraphics[page=2,width=0.5\textwidth]{Analysis/Centre/RGDP_11/Ctr_RGDP_fit.pdf}}

  \caption{Inferred density of the radially symmetric profiles shown by means of the MAP value (red line) and 100 samples from the posterior distribution (grey lines). 
  For comparison the data has been binned with Poissonian uncertainties (black dots).}
\label{fig:PSDctr}
\end {figure*}

\begin{figure*}
  \includegraphics[page=4,width=\textwidth]{Analysis/Centre/EFF_11/Ctr_EFF_fit.pdf}
  \caption{Projections of the posterior distribution for the radially symmetric EFF model.}
\label{fig:EFFctr_posterior}
\end {figure*}

\begin{figure*}
 \centering
   \includegraphics[page=4,width=\textwidth]{Analysis/Centre/GDP_11/Ctr_GDP_fit.pdf}
  \caption{Projections of the posterior distribution for the radially symmetric GDP model.}
\label{fig:GDPctr_posterior}
\end {figure*}

\begin {figure*}
 \centering
   \includegraphics[page=4,width=\textwidth]{Analysis/Centre/GKing_11/Ctr_GKing_fit.pdf}
  \caption{Projections of the posterior distribution for the radially symmetric GKing model.}
\label{fig:GKingctr}
\end {figure*}

\begin {figure*}
 \centering
   \includegraphics[page=4,width=\textwidth]{Analysis/Centre/King_11/Ctr_King_fit.pdf}
  \caption{Projections of the posterior distribution for the radially symmetric King's model .}
\label{fig:Kingctr}
\end {figure*}

\begin {figure*}
 \centering
   \includegraphics[page=4,width=\textwidth]{Analysis/Centre/OGKing_11/Ctr_OGKing_fit.pdf}
  \caption{Projections of the posterior distribution for the radially symmetric OGKing model.}
\label{fig:OGKingctr}
\end {figure*}

\begin {figure*}
   \includegraphics[page=4,width=\textwidth]{Analysis/Centre/RGDP_11/Ctr_RGDP_fit.pdf}
  \caption{Projections of the posterior distribution for the radially symmetric RGDP model.}
\label{fig:RGDPctr}
\end {figure*}

\begin{table*}
  \centering
      \caption[]{Covariance matrix of the biaxially symmetric EFF model.}
  \begin{tabular}{lrrrrrr}
\toprule
{} &  $\alpha_c\ \ [^\circ]$ &  $\delta_c\ \ [^\circ]$ &  $\phi$ [radians] &  $r_{ca}$ [pc] &  $r_{cb}$ [pc] &  $\gamma$ \\
\midrule
$\alpha_c\ \ [^\circ]$ &                   0.007 &                  -0.000 &            -0.001 &          0.000 &          0.000 &     0.000 \\
$\delta_c\ \ [^\circ]$ &                  -0.000 &                   0.006 &             0.001 &          0.000 &          0.001 &     0.001 \\
$\phi$ [radians]       &                  -0.001 &                   0.001 &             0.063 &          0.000 &         -0.002 &    -0.000 \\
$r_{ca}$ [pc]          &                   0.000 &                   0.000 &             0.000 &          0.131 &          0.056 &     0.049 \\
$r_{cb}$ [pc]          &                   0.000 &                   0.001 &            -0.002 &          0.056 &          0.093 &     0.047 \\
$\gamma$               &                   0.000 &                   0.001 &            -0.000 &          0.049 &          0.047 &     0.040 \\
\bottomrule
\end{tabular}

   \end{table*}
   
\begin{table*}
  \centering
      \caption[]{Covariance matrix of the biaxially symmetric GDP model.}
  \begin{tabular}{lrrrrrrrr}
\toprule
{} &  $\alpha_c\ \ [^\circ]$ &  $\delta_c\ \ [^\circ]$ &  $\phi$ [radians] &  $r_{ca}$ [pc] &  $r_{cb}$ [pc] &  $\alpha$ &  $\beta$ &  $\gamma$ \\
\midrule
$\alpha_c\ \ [^\circ]$ &                   0.007 &                  -0.001 &            -0.001 &          0.002 &          0.000 &    -0.001 &    0.000 &     0.001 \\
$\delta_c\ \ [^\circ]$ &                  -0.001 &                   0.005 &             0.001 &          0.005 &          0.005 &     0.002 &    0.005 &    -0.001 \\
$\phi$ [radians]       &                  -0.001 &                   0.001 &             0.185 &          0.051 &          0.031 &     0.008 &    0.043 &    -0.010 \\
$r_{ca}$ [pc]          &                   0.002 &                   0.005 &             0.051 &          1.204 &          0.801 &     0.101 &    0.547 &    -0.006 \\
$r_{cb}$ [pc]          &                   0.000 &                   0.005 &             0.031 &          0.801 &          0.788 &     0.074 &    0.480 &    -0.009 \\
$\alpha$               &                  -0.001 &                   0.002 &             0.008 &          0.101 &          0.074 &     0.044 &    0.070 &    -0.016 \\
$\beta$                &                   0.000 &                   0.005 &             0.043 &          0.547 &          0.480 &     0.070 &    0.363 &    -0.033 \\
$\gamma$               &                   0.001 &                  -0.001 &            -0.010 &         -0.006 &         -0.009 &    -0.016 &   -0.033 &     0.025 \\
\bottomrule
\end{tabular}

   \end{table*}
   
 \begin{table*}
  \centering
      \caption[]{Covariance matrix of the biaxially symmetric GKing model.}
  \begin{tabular}{lrrrrrrrrr}
\toprule
{} &  $\alpha_c\ \ [^\circ]$ &  $\delta_c\ \ [^\circ]$ &  $\phi$ [radians] &  $r_{ca}$ [pc] &  $r_{ta}$ [pc] &  $r_{cb}$ [pc] &  $r_{tb}$ [pc] &  $\alpha$ &  $\beta$ \\
\midrule
$\alpha_c\ \ [^\circ]$ &                   0.013 &                  -0.001 &            -0.004 &          0.001 &         -0.004 &         -0.002 &         -0.015 &     0.002 &   -0.001 \\
$\delta_c\ \ [^\circ]$ &                  -0.001 &                   0.010 &             0.001 &         -0.002 &          0.014 &          0.004 &          0.005 &    -0.002 &    0.002 \\
$\phi$ [radians]       &                  -0.004 &                   0.001 &             0.256 &         -0.029 &          0.205 &          0.058 &         -0.003 &    -0.022 &    0.027 \\
$r_{ca}$ [pc]          &                   0.001 &                  -0.002 &            -0.029 &          1.587 &         -0.041 &          0.379 &          0.437 &     0.194 &    0.131 \\
$r_{ta}$ [pc]          &                  -0.004 &                   0.014 &             0.205 &         -0.041 &         25.850 &          0.956 &          5.304 &    -0.227 &    0.677 \\
$r_{cb}$ [pc]          &                  -0.002 &                   0.004 &             0.058 &          0.379 &          0.956 &          0.433 &          0.575 &     0.031 &    0.144 \\
$r_{tb}$ [pc]          &                  -0.015 &                   0.005 &            -0.003 &          0.437 &          5.304 &          0.575 &          6.150 &     0.015 &    0.436 \\
$\alpha$               &                   0.002 &                  -0.002 &            -0.022 &          0.194 &         -0.227 &          0.031 &          0.015 &     0.291 &    0.028 \\
$\beta$                &                  -0.001 &                   0.002 &             0.027 &          0.131 &          0.677 &          0.144 &          0.436 &     0.028 &    0.076 \\
\bottomrule
\end{tabular}

   \end{table*}
   
 \begin{table*}
  \centering
      \caption[]{Covariance matrix of the biaxially symmetric King model.}
  \begin{tabular}{lrrrrrrr}
\toprule
{} &  $\alpha_c\ \ [^\circ]$ &  $\delta_c\ \ [^\circ]$ &  $\phi$ [radians] &  $r_{ca}$ [pc] &  $r_{ta}$ [pc] &  $r_{cb}$ [pc] &  $r_{tb}$ [pc] \\
\midrule
$\alpha_c\ \ [^\circ]$ &                   0.019 &                  -0.000 &            -0.004 &          0.002 &         -0.057 &          0.002 &          0.001 \\
$\delta_c\ \ [^\circ]$ &                  -0.000 &                   0.015 &             0.007 &         -0.010 &          0.094 &          0.001 &         -0.001 \\
$\phi$ [radians]       &                  -0.004 &                   0.007 &             0.359 &         -0.124 &          1.582 &          0.010 &         -0.190 \\
$r_{ca}$ [pc]          &                   0.002 &                  -0.010 &            -0.124 &          1.428 &         -6.175 &          0.074 &         -1.548 \\
$r_{ta}$ [pc]          &                  -0.057 &                   0.094 &             1.582 &         -6.175 &        371.812 &         -1.108 &         23.841 \\
$r_{cb}$ [pc]          &                   0.002 &                   0.001 &             0.010 &          0.074 &         -1.108 &          0.154 &         -1.157 \\
$r_{tb}$ [pc]          &                   0.001 &                  -0.001 &            -0.190 &         -1.548 &         23.841 &         -1.157 &         53.033 \\
\bottomrule
\end{tabular}

   \end{table*}
   
 \begin{table*}
  \centering
      \caption[]{Covariance matrix of the biaxially symmetric OGKing model.}
  \begin{tabular}{lrrrrrrr}
\toprule
{} &  $\alpha_c\ \ [^\circ]$ &  $\delta_c\ \ [^\circ]$ &  $\phi$ [radians] &  $r_{ca}$ [pc] &  $r_{ta}$ [pc] &  $r_{cb}$ [pc] &  $r_{tb}$ [pc] \\
\midrule
$\alpha_c\ \ [^\circ]$ &                   0.009 &                  -0.001 &            -0.000 &         -0.001 &         -0.006 &         -0.001 &         -0.002 \\
$\delta_c\ \ [^\circ]$ &                  -0.001 &                   0.007 &             0.002 &         -0.004 &          0.011 &          0.001 &         -0.002 \\
$\phi$ [radians]       &                  -0.000 &                   0.002 &             0.194 &         -0.031 &          0.166 &          0.009 &         -0.121 \\
$r_{ca}$ [pc]          &                  -0.001 &                  -0.004 &            -0.031 &          0.248 &         -0.462 &          0.011 &         -0.080 \\
$r_{ta}$ [pc]          &                  -0.006 &                   0.011 &             0.166 &         -0.462 &         14.223 &         -0.095 &         -0.391 \\
$r_{cb}$ [pc]          &                  -0.001 &                   0.001 &             0.009 &          0.011 &         -0.095 &          0.059 &         -0.176 \\
$r_{tb}$ [pc]          &                  -0.002 &                  -0.002 &            -0.121 &         -0.080 &         -0.391 &         -0.176 &          3.509 \\
\bottomrule
\end{tabular}

   \end{table*}
   
 \begin{table*}
  \centering
      \caption[]{Covariance matrix of the biaxially symmetric RGDP model.}
  \begin{tabular}{lrrrrrrr}
\toprule
{} &  $\alpha_c\ \ [^\circ]$ &  $\delta_c\ \ [^\circ]$ &  $\phi$ [radians] &  $r_{ca}$ [pc] &  $r_{cb}$ [pc] &  $\alpha$ &  $\beta$ \\
\midrule
$\alpha_c\ \ [^\circ]$ &                   0.010 &                  -0.001 &            -0.003 &          0.003 &         -0.000 &    -0.000 &   -0.000 \\
$\delta_c\ \ [^\circ]$ &                  -0.001 &                   0.008 &             0.001 &          0.006 &          0.005 &     0.001 &    0.004 \\
$\phi$ [radians]       &                  -0.003 &                   0.001 &             0.208 &          0.037 &          0.030 &     0.001 &    0.029 \\
$r_{ca}$ [pc]          &                   0.003 &                   0.006 &             0.037 &          1.320 &          0.864 &     0.115 &    0.590 \\
$r_{cb}$ [pc]          &                  -0.000 &                   0.005 &             0.030 &          0.864 &          0.823 &     0.084 &    0.507 \\
$\alpha$               &                  -0.000 &                   0.001 &             0.001 &          0.115 &          0.084 &     0.045 &    0.062 \\
$\beta$                &                  -0.000 &                   0.004 &             0.029 &          0.590 &          0.507 &     0.062 &    0.355 \\
\bottomrule
\end{tabular}

   \end{table*}

\begin{figure*}
\subfloat[][EFF]{   \includegraphics[page=2,width=0.5\textwidth]{Analysis/Elliptic/EFF_11/Ell_EFF_fit.pdf}}
\subfloat[][GDP]{   \includegraphics[page=2,width=0.5\textwidth]{Analysis/Elliptic/GDP_11/Ell_GDP_fit.pdf}}
\\
\subfloat[][GKing]{ \includegraphics[page=2,width=0.5\textwidth]{Analysis/Elliptic/GKing_11/Ell_GKing_fit.pdf}}
\subfloat[][King]{  \includegraphics[page=2,width=0.5\textwidth]{Analysis/Elliptic/King_11/Ell_King_fit.pdf}}
\\
\subfloat[][OGKing]{\includegraphics[page=2,width=0.5\textwidth]{Analysis/Elliptic/OGKing_11/Ell_OGKing_fit.pdf}}
\subfloat[][RGDP]{  \includegraphics[page=2,width=0.5\textwidth]{Analysis/Elliptic/RGDP_11/Ell_RGDP_fit.pdf}}

  \caption{Inferred density of the biaxially symmetric profiles shown by means of the MAP value (red line) and 100 samples from the posterior distribution (grey lines). 
  For comparison the data has been binned with Poissonian uncertainties (black dots).}
\label{fig:PSDEll}
\end {figure*}

\begin {figure*}
 \centering
   \includegraphics[page=4,width=\textwidth]{Analysis/Elliptic/EFF_11/Ell_EFF_fit.pdf}
  \caption{Projections of the posterior distribution for the biaxially symmetric EFF  model .}
\label{fig:EFFEll}
\end {figure*}
   
\begin {figure*}
 \centering
   \includegraphics[page=4,width=\textwidth]{Analysis/Elliptic/GDP_11/Ell_GDP_fit.pdf}
  \caption{Projections of the posterior distribution for the biaxially symmetric GDP  model .}
\label{fig:GDPEll}
\end {figure*}

\begin {figure*}
 \centering
   \includegraphics[page=4,width=\textwidth]{Analysis/Elliptic/GKing_11/Ell_GKing_fit.pdf}
  \caption{Projections of the posterior distribution for the biaxially symmetric GKing  model.}
\label{fig:GKingEll}
\end {figure*}

\begin {figure*}
 \centering
   \includegraphics[page=4,width=\textwidth]{Analysis/Elliptic/King_11/Ell_King_fit.pdf}
  \caption{Projections of the posterior distribution for the biaxially symmetric King's  model.}
\label{fig:KingEll}
\end {figure*}

\begin {figure*}
 \centering
   \includegraphics[page=4,width=\textwidth]{Analysis/Elliptic/OGKing_11/Ell_OGKing_fit.pdf}
  \caption{Projections of the posterior distribution for the biaxially symmetric OGKing model.}
\label{fig:OGKingEll}
\end {figure*}

\begin {figure*}
 \centering
   \includegraphics[page=4,width=\textwidth]{Analysis/Elliptic/RGDP_11/Ell_RGDP_fit.pdf}
  \caption{Projections of the posterior distribution for the biaxially symmetric RGDP model.}
\label{fig:RGDPEll}
\end {figure*}

\begin{figure*}
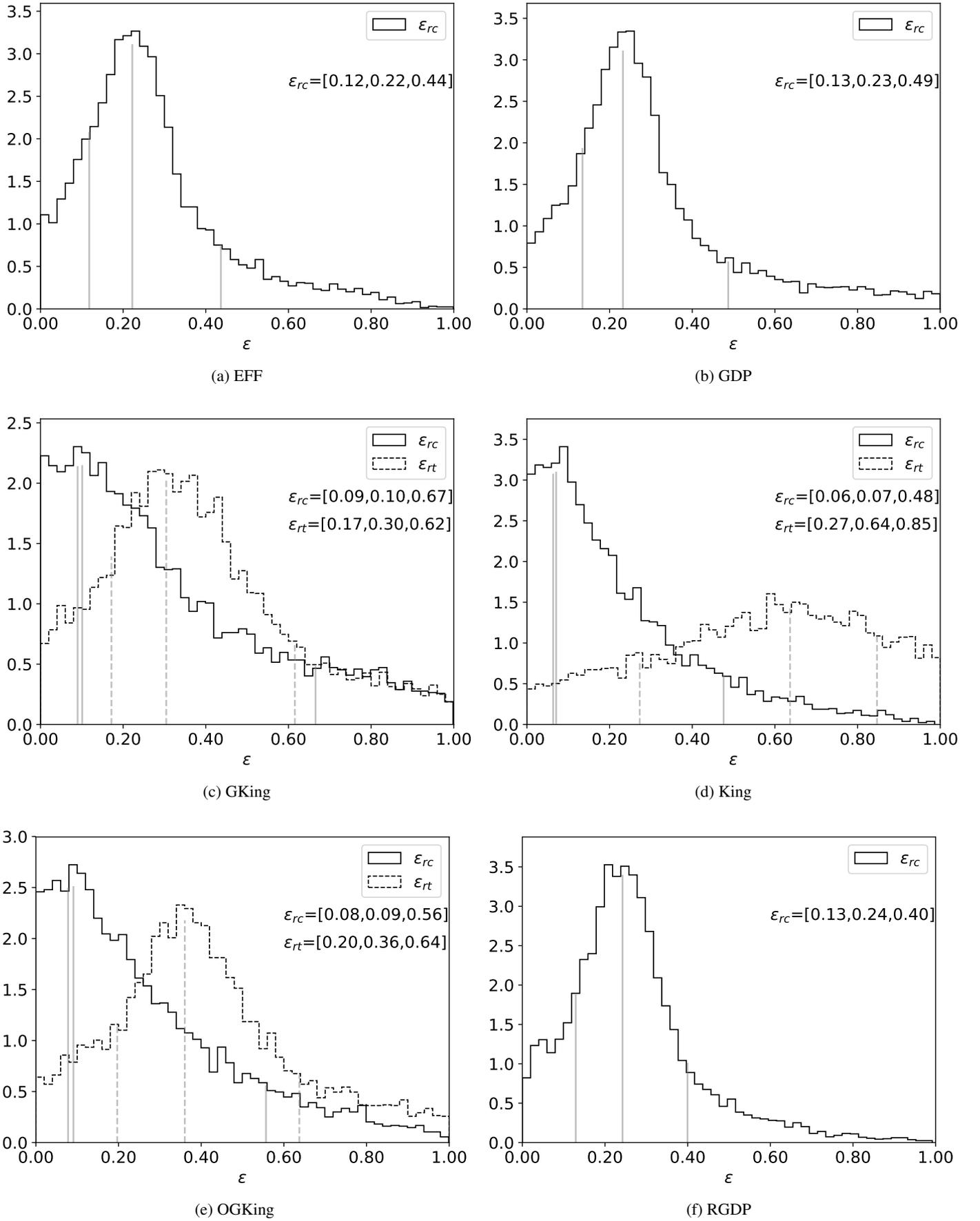
  
\subfloat[][EFF]{   \includegraphics[page=5,width=0.5\textwidth]{Analysis/Elliptic/EFF_11/Ell_EFF_fit.pdf}}
\subfloat[][GDP]{   \includegraphics[page=5,width=0.5\textwidth]{Analysis/Elliptic/GDP_11/Ell_GDP_fit.pdf}}
\\
\subfloat[][GKing]{ \includegraphics[page=5,width=0.5\textwidth]{Analysis/Elliptic/GKing_11/Ell_GKing_fit.pdf}}
\subfloat[][King]{  \includegraphics[page=5,width=0.5\textwidth]{Analysis/Elliptic/King_11/Ell_King_fit.pdf}}
\\
\subfloat[][OGKing]{\includegraphics[page=5,width=0.5\textwidth]{Analysis/Elliptic/OGKing_11/Ell_OGKing_fit.pdf}}
\subfloat[][RGDP]{  \includegraphics[page=5,width=0.5\textwidth]{Analysis/Elliptic/RGDP_11/Ell_RGDP_fit.pdf}}
  \caption{Ellipticity distributions of the biaxially symmetric models. The numbers shown in brackets represent the 16th percentile, the mode, and the 84th percentile (also shown by means of vertical grey lines).}
\label{fig:EllipticityEll}
\end {figure*}


 \begin{table*}
  \centering
      \caption[]{Covariance matrix of the luminosity segregated EFF model.}
  \begin{tabular}{lrrrrrrr}
\toprule
{} &  $\alpha_c\ \ [^\circ]$ &  $\delta_c\ \ [^\circ]$ &  $\phi$ [radians] &  $r_{ca}$ [pc] &  $r_{cb}$ [pc] &  $\gamma$ &  $\kappa$ [pc$\cdot \rm{mag}^{-1}$] \\
\midrule
$\alpha_c\ \ [^\circ]$             &                   0.007 &                  -0.001 &            -0.000 &          0.001 &         -0.001 &    -0.000 &                              -0.000 \\
$\delta_c\ \ [^\circ]$             &                  -0.001 &                   0.006 &             0.002 &          0.001 &          0.002 &     0.001 &                               0.001 \\
$\phi$ [radians]                   &                  -0.000 &                   0.002 &             0.085 &          0.009 &          0.002 &     0.005 &                               0.001 \\
$r_{ca}$ [pc]                      &                   0.001 &                   0.001 &             0.009 &          0.151 &          0.078 &     0.060 &                               0.008 \\
$r_{cb}$ [pc]                      &                  -0.001 &                   0.002 &             0.002 &          0.078 &          0.126 &     0.061 &                               0.012 \\
$\gamma$                           &                  -0.000 &                   0.001 &             0.005 &          0.060 &          0.061 &     0.048 &                               0.004 \\
$\kappa$ [pc$\cdot \rm{mag}^{-1}$] &                  -0.000 &                   0.001 &             0.001 &          0.008 &          0.012 &     0.004 &                               0.006 \\
\bottomrule
\end{tabular}

   \end{table*}

 \begin{table*}
  \centering
      \caption[]{Covariance matrix of the luminosity segregated GDP model.}
  \begin{tabular}{lrrrrrrrrr}
\toprule
{} &  $\alpha_c\ \ [^\circ]$ &  $\delta_c\ \ [^\circ]$ &  $\phi$ [radians] &  $r_{ca}$ [pc] &  $r_{cb}$ [pc] &  $\alpha$ &  $\beta$ &  $\gamma$ &  $\kappa$ [pc$\cdot \rm{mag}^{-1}$] \\
\midrule
$\alpha_c\ \ [^\circ]$             &                   0.010 &                  -0.001 &             0.000 &         -0.000 &         -0.003 &    -0.001 &   -0.003 &     0.001 &                              -0.001 \\
$\delta_c\ \ [^\circ]$             &                  -0.001 &                   0.008 &             0.002 &          0.003 &          0.004 &     0.000 &    0.004 &    -0.001 &                               0.001 \\
$\phi$ [radians]                   &                   0.000 &                   0.002 &             0.321 &          0.102 &          0.064 &     0.010 &    0.071 &    -0.012 &                               0.014 \\
$r_{ca}$ [pc]                      &                  -0.000 &                   0.003 &             0.102 &          1.122 &          0.789 &     0.092 &    0.506 &    -0.003 &                               0.061 \\
$r_{cb}$ [pc]                      &                  -0.003 &                   0.004 &             0.064 &          0.789 &          0.828 &     0.072 &    0.472 &    -0.011 &                               0.069 \\
$\alpha$                           &                  -0.001 &                   0.000 &             0.010 &          0.092 &          0.072 &     0.051 &    0.068 &    -0.019 &                               0.004 \\
$\beta$                            &                  -0.003 &                   0.004 &             0.071 &          0.506 &          0.472 &     0.068 &    0.354 &    -0.040 &                               0.040 \\
$\gamma$                           &                   0.001 &                  -0.001 &            -0.012 &         -0.003 &         -0.011 &    -0.019 &   -0.040 &     0.031 &                              -0.003 \\
$\kappa$ [pc$\cdot \rm{mag}^{-1}$] &                  -0.001 &                   0.001 &             0.014 &          0.061 &          0.069 &     0.004 &    0.040 &    -0.003 &                               0.016 \\
\bottomrule
\end{tabular}

   \end{table*}
 \begin{table*}
  \centering
      \caption[]{Covariance matrix of the luminosity segregated GKing model.}
  \begin{tabular}{lrrrrrrrrrr}
\toprule
{} &  $\alpha_c\ \ [^\circ]$ &  $\delta_c\ \ [^\circ]$ &  $\phi$ [radians] &  $r_{ca}$ [pc] &  $r_{ta}$ [pc] &  $r_{cb}$ [pc] &  $r_{tb}$ [pc] &  $\alpha$ &  $\beta$ &  $\kappa$ [pc$\cdot \rm{mag}^{-1}$] \\
\midrule
$\alpha_c\ \ [^\circ]$             &                   0.015 &                  -0.001 &            -0.003 &          0.002 &         -0.054 &         -0.007 &         -0.032 &     0.003 &   -0.003 &                              -0.002 \\
$\delta_c\ \ [^\circ]$             &                  -0.001 &                   0.012 &             0.006 &          0.004 &          0.076 &          0.017 &          0.033 &    -0.000 &    0.007 &                               0.003 \\
$\phi$ [radians]                   &                  -0.003 &                   0.006 &             0.306 &          0.060 &          0.406 &          0.186 &          0.138 &    -0.017 &    0.067 &                               0.027 \\
$r_{ca}$ [pc]                      &                   0.002 &                   0.004 &             0.060 &          4.645 &          1.703 &          2.322 &          1.919 &     0.388 &    0.563 &                               0.167 \\
$r_{ta}$ [pc]                      &                  -0.054 &                   0.076 &             0.406 &          1.703 &         85.273 &          4.036 &         19.497 &    -0.637 &    2.217 &                               0.511 \\
$r_{cb}$ [pc]                      &                  -0.007 &                   0.017 &             0.186 &          2.322 &          4.036 &          2.409 &          2.465 &     0.135 &    0.632 &                               0.226 \\
$r_{tb}$ [pc]                      &                  -0.032 &                   0.033 &             0.138 &          1.919 &         19.497 &          2.465 &         17.347 &    -0.149 &    1.308 &                               0.292 \\
$\alpha$                           &                   0.003 &                  -0.000 &            -0.017 &          0.388 &         -0.637 &          0.135 &         -0.149 &     0.342 &    0.023 &                              -0.003 \\
$\beta$                            &                  -0.003 &                   0.007 &             0.067 &          0.563 &          2.217 &          0.632 &          1.308 &     0.023 &    0.232 &                               0.067 \\
$\kappa$ [pc$\cdot \rm{mag}^{-1}$] &                  -0.002 &                   0.003 &             0.027 &          0.167 &          0.511 &          0.226 &          0.292 &    -0.003 &    0.067 &                               0.037 \\
\bottomrule
\end{tabular}

   \end{table*}
 \begin{table*}
  \centering
      \caption[]{Covariance matrix of the luminosity segregated King model.}
  \begin{tabular}{lrrrrrrrr}
\toprule
{} &  $\alpha_c\ \ [^\circ]$ &  $\delta_c\ \ [^\circ]$ &  $\phi$ [radians] &  $r_{ca}$ [pc] &  $r_{ta}$ [pc] &  $r_{cb}$ [pc] &  $r_{tb}$ [pc] &  $\kappa$ [pc$\cdot \rm{mag}^{-1}$] \\
\midrule
$\alpha_c\ \ [^\circ]$             &                   0.018 &                  -0.001 &            -0.003 &          0.015 &         -0.105 &          0.002 &         -0.019 &                              -0.000 \\
$\delta_c\ \ [^\circ]$             &                  -0.001 &                   0.014 &             0.004 &         -0.009 &          0.123 &         -0.001 &          0.010 &                               0.000 \\
$\phi$ [radians]                   &                  -0.003 &                   0.004 &             0.328 &         -0.118 &          1.317 &         -0.005 &         -0.064 &                               0.004 \\
$r_{ca}$ [pc]                      &                   0.015 &                  -0.009 &            -0.118 &          1.547 &         -5.348 &          0.110 &         -1.293 &                               0.002 \\
$r_{ta}$ [pc]                      &                  -0.105 &                   0.123 &             1.317 &         -5.348 &        220.118 &         -1.423 &         16.847 &                              -0.041 \\
$r_{cb}$ [pc]                      &                   0.002 &                  -0.001 &            -0.005 &          0.110 &         -1.423 &          0.192 &         -1.046 &                               0.014 \\
$r_{tb}$ [pc]                      &                  -0.019 &                   0.010 &            -0.064 &         -1.293 &         16.847 &         -1.046 &         32.386 &                              -0.069 \\
$\kappa$ [pc$\cdot \rm{mag}^{-1}$] &                  -0.000 &                   0.000 &             0.004 &          0.002 &         -0.041 &          0.014 &         -0.069 &                               0.004 \\
\bottomrule
\end{tabular}

   \end{table*}
 \begin{table*}
  \centering
      \caption[]{Covariance matrix of the luminosity segregated OGKing model.}

  \begin{tabular}{lrrrrrrrr}
\toprule
{} &  $\alpha_c\ \ [^\circ]$ &  $\delta_c\ \ [^\circ]$ &  $\phi$ [radians] &  $r_{ca}$ [pc] &  $r_{ta}$ [pc] &  $r_{cb}$ [pc] &  $r_{tb}$ [pc] &  $\kappa$ [pc$\cdot \rm{mag}^{-1}$] \\
\midrule
$\alpha_c\ \ [^\circ]$             &                   0.009 &                  -0.000 &            -0.002 &          0.002 &         -0.010 &         -0.000 &         -0.005 &                              -0.001 \\
$\delta_c\ \ [^\circ]$             &                  -0.000 &                   0.006 &             0.002 &         -0.002 &          0.013 &          0.002 &         -0.005 &                               0.001 \\
$\phi$ [radians]                   &                  -0.002 &                   0.002 &             0.176 &         -0.019 &          0.137 &          0.008 &         -0.104 &                               0.008 \\
$r_{ca}$ [pc]                      &                   0.002 &                  -0.002 &            -0.019 &          0.175 &         -0.381 &          0.025 &         -0.080 &                              -0.002 \\
$r_{ta}$ [pc]                      &                  -0.010 &                   0.013 &             0.137 &         -0.381 &         10.973 &         -0.093 &         -0.386 &                               0.009 \\
$r_{cb}$ [pc]                      &                  -0.000 &                   0.002 &             0.008 &          0.025 &         -0.093 &          0.070 &         -0.186 &                               0.008 \\
$r_{tb}$ [pc]                      &                  -0.005 &                  -0.005 &            -0.104 &         -0.080 &         -0.386 &         -0.186 &          2.807 &                              -0.023 \\
$\kappa$ [pc$\cdot \rm{mag}^{-1}$] &                  -0.001 &                   0.001 &             0.008 &         -0.002 &          0.009 &          0.008 &         -0.023 &                               0.005 \\
\bottomrule
\end{tabular}

   \end{table*}
 \begin{table*}
  \centering
      \caption[]{Covariance matrix of the luminosity segregated RGDP model.}
  \begin{tabular}{lrrrrrrrr}
\toprule
{} &  $\alpha_c\ \ [^\circ]$ &  $\delta_c\ \ [^\circ]$ &  $\phi$ [radians] &  $r_{ca}$ [pc] &  $r_{cb}$ [pc] &  $\alpha$ &  $\beta$ &  $\kappa$ [pc$\cdot \rm{mag}^{-1}$] \\
\midrule
$\alpha_c\ \ [^\circ]$             &                   0.010 &                  -0.000 &            -0.000 &          0.002 &          0.001 &    -0.000 &    0.001 &                              -0.001 \\
$\delta_c\ \ [^\circ]$             &                  -0.000 &                   0.008 &             0.003 &          0.006 &          0.009 &     0.001 &    0.006 &                               0.002 \\
$\phi$ [radians]                   &                  -0.000 &                   0.003 &             0.237 &          0.078 &          0.056 &     0.003 &    0.046 &                               0.012 \\
$r_{ca}$ [pc]                      &                   0.002 &                   0.006 &             0.078 &          1.481 &          1.074 &     0.125 &    0.667 &                               0.081 \\
$r_{cb}$ [pc]                      &                   0.001 &                   0.009 &             0.056 &          1.074 &          1.078 &     0.090 &    0.609 &                               0.088 \\
$\alpha$                           &                  -0.000 &                   0.001 &             0.003 &          0.125 &          0.090 &     0.050 &    0.062 &                               0.005 \\
$\beta$                            &                   0.001 &                   0.006 &             0.046 &          0.667 &          0.609 &     0.062 &    0.390 &                               0.047 \\
$\kappa$ [pc$\cdot \rm{mag}^{-1}$] &                  -0.001 &                   0.002 &             0.012 &          0.081 &          0.088 &     0.005 &    0.047 &                               0.016 \\
\bottomrule
\end{tabular}

   \end{table*}
   
 \begin{figure*}
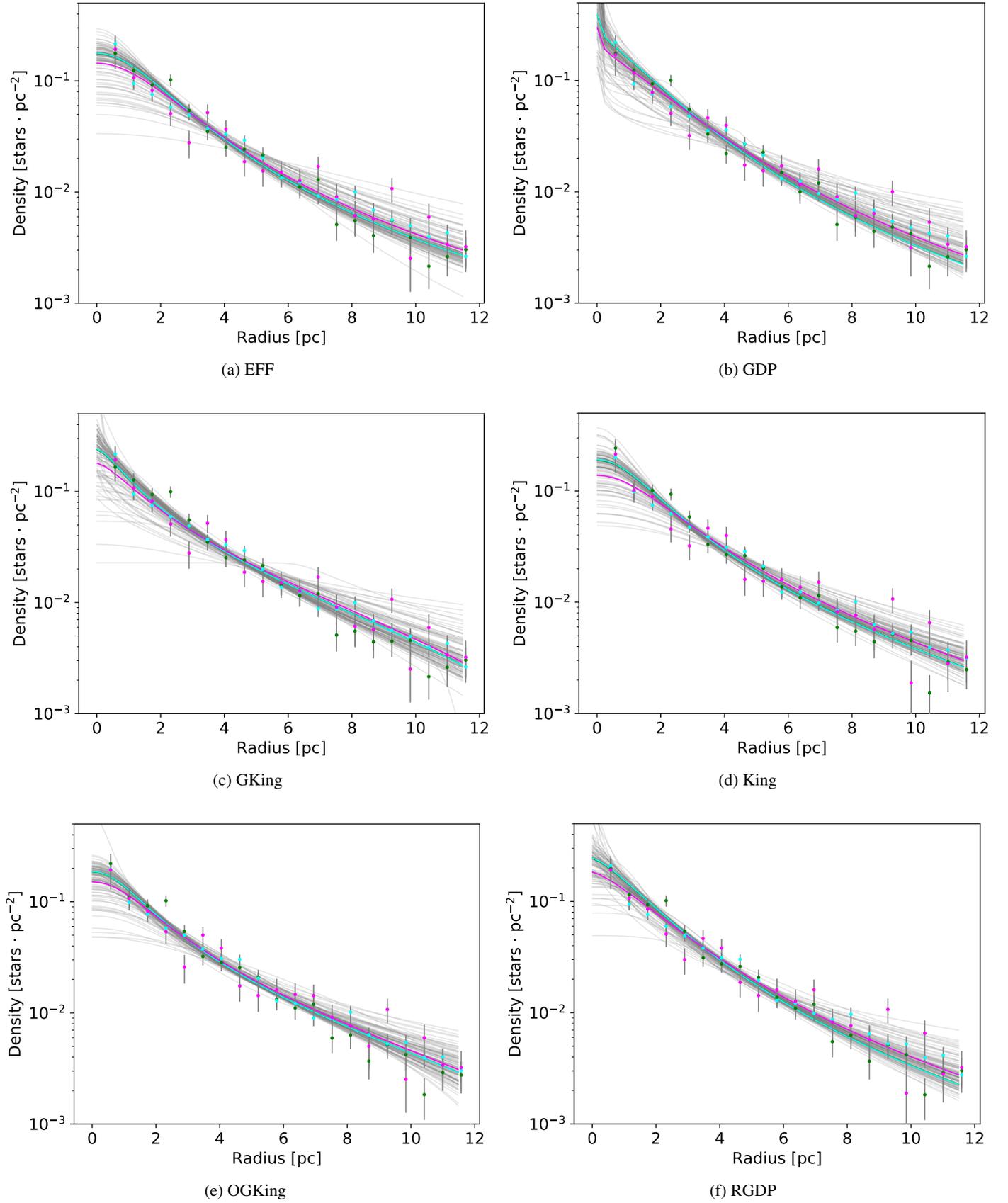

\subfloat[][EFF]{   \includegraphics[page=5,width=0.5\textwidth]{Analysis/Segregated/EFF_11/Seg_EFF_fit.pdf}}
\subfloat[][GDP]{   \includegraphics[page=5,width=0.5\textwidth]{Analysis/Segregated/GDP_11/Seg_GDP_fit.pdf}}
\\
\subfloat[][GKing]{ \includegraphics[page=5,width=0.5\textwidth]{Analysis/Segregated/GKing_11/Seg_GKing_fit.pdf}}
\subfloat[][King]{  \includegraphics[page=5,width=0.5\textwidth]{Analysis/Segregated/King_11/Seg_King_fit.pdf}}
\\
\subfloat[][OGKing]{\includegraphics[page=5,width=0.5\textwidth]{Analysis/Segregated/OGKing_11/Seg_OGKing_fit.pdf}}
\subfloat[][RGDP]{  \includegraphics[page=5,width=0.5\textwidth]{Analysis/Segregated/RGDP_11/Seg_RGDP_fit.pdf}}

  \caption{Inferred density of the luminosity segregated models.  The data are binned in three bins of the J band: $J < 12$, $12 \lesssim J \lesssim 15$, and $15 < J$ (with colours green, cyan and magenta, respectively). The MAP is shown by means of three coloured solid lines, the colours correspond to those of the $J$ band bins. In these MAPs, the core radius is increased accordingly to Eq. \ref{eq:segregation} using the mean value of the J band in each bin.
Also shown are 100 samples from the posterior distribution (grey lines).}
\label{fig:PSDSeg}
\end {figure*}

\begin {figure*}
 \centering
   \includegraphics[page=4,width=\textwidth]{Analysis/Segregated/EFF_11/Seg_EFF_fit.pdf}
  \caption{Projections of the posterior distribution for the luminosity segregated EFF model.}
\label{fig:EFFSeg}
\end {figure*}

\begin {figure*}
 \centering
   \includegraphics[page=4,width=\textwidth]{Analysis/Segregated/GDP_11/Seg_GDP_fit.pdf}
  \caption{Projections of the posterior distribution for the luminosity segregated GDP model.}
\label{fig:GDPSeg}
\end {figure*}

\begin {figure*}
 \centering
   \includegraphics[page=4,width=\textwidth]{Analysis/Segregated/GKing_11/Seg_GKing_fit.pdf}
  \caption{Projections of the posterior distribution for the luminosity segregated GKing model.}
\label{fig:GKingSeg}
\end {figure*}

\begin {figure*}
 \centering
   \includegraphics[page=4,width=\textwidth]{Analysis/Segregated/King_11/Seg_King_fit.pdf}
  \caption{Projections of the posterior distribution for the luminosity segregated King's model.}
\label{fig:KingSeg}
\end {figure*}

\begin {figure*}
 \centering
   \includegraphics[page=4,width=\textwidth]{Analysis/Segregated/OGKing_11/Seg_OGKing_fit.pdf}
  \caption{Projections of the posterior distribution for the luminosity segregated OGKing model.}
\label{fig:OGKingSeg}
\end {figure*}

\begin {figure*}
 \centering
   \includegraphics[page=4,width=\textwidth]{Analysis/Segregated/RGDP_11/Seg_RGDP_fit.pdf}
  \caption{Projections of the posterior distribution for the luminosity segregated RGDP model.}
\label{fig:RGDPSeg}
\end {figure*}

\begin{figure*}
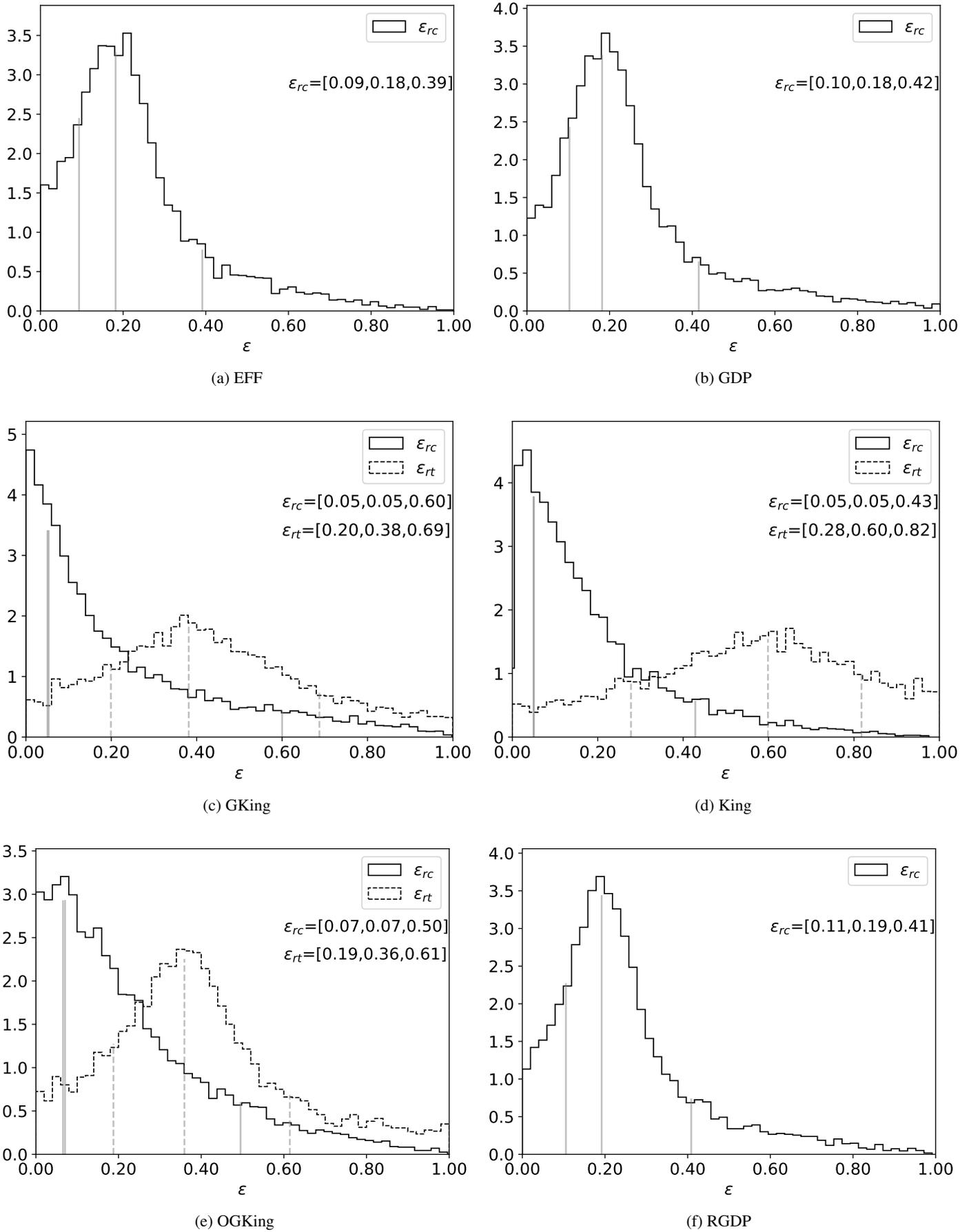

\subfloat[][EFF]{   \includegraphics[page=6,width=0.5\textwidth]{Analysis/Segregated/EFF_11/Seg_EFF_fit.pdf}}
\subfloat[][GDP]{   \includegraphics[page=6,width=0.5\textwidth]{Analysis/Segregated/GDP_11/Seg_GDP_fit.pdf}}
\\
\subfloat[][GKing]{ \includegraphics[page=6,width=0.5\textwidth]{Analysis/Segregated/GKing_11/Seg_GKing_fit.pdf}}
\subfloat[][King]{  \includegraphics[page=6,width=0.5\textwidth]{Analysis/Segregated/King_11/Seg_King_fit.pdf}}
\\
\subfloat[][OGKing]{\includegraphics[page=6,width=0.5\textwidth]{Analysis/Segregated/OGKing_11/Seg_OGKing_fit.pdf}}
\subfloat[][RGDP]{  \includegraphics[page=6,width=0.5\textwidth]{Analysis/Segregated/RGDP_11/Seg_RGDP_fit.pdf}}

  \caption{Ellipticity distributions of the luminosity segregated models. The numbers shown in brackets represent the 16th percentile, the mode, and the 84th percentile (also shown by means of vertical grey lines).}
\label{fig:EllipticitySeg}
\end {figure*}

\begin{figure}
\begin{center}
  \includegraphics[page=1,width=\columnwidth]{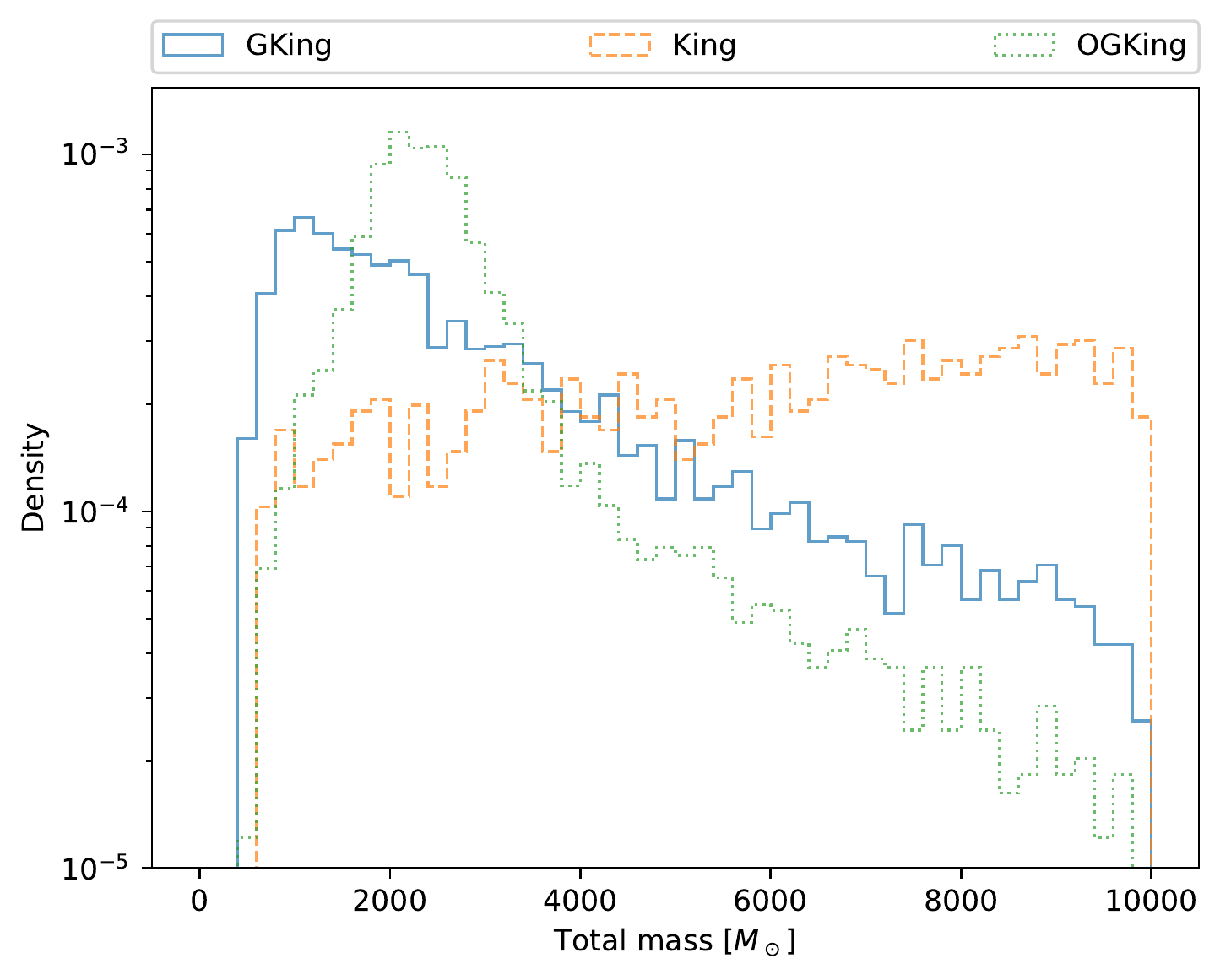}
\caption{Distribution of the total mass of the cluster derived from each radially symmetric model of the King's family.}
\label{fig:RadMass}
\end{center}
\end{figure}

\begin{figure}
\begin{center}
  \includegraphics[page=2,width=\columnwidth]{Analysis/BayesFactors/Msys_11.pdf}
\caption{Distribution of the total mass of the cluster derived from each biaxially symmetric model of the King's family.}
\label{fig:EllMass}
\end{center}
\end{figure}

\end{appendix}
   
\end{document}